\documentclass[9pt]{article}

\usepackage[latin1]{inputenc}
\usepackage[english]{babel}

\usepackage{amsmath,amssymb,stmaryrd,amsthm}
\usepackage{wasysym}
\usepackage{subeqnarray}
\usepackage{verbatim}
\usepackage[colorlinks=false,linktocpage]{hyperref}%
\usepackage{multicol}
\usepackage[bottom]{footmisc}
\usepackage{cancel}
\usepackage{units}
\usepackage{pgf}
\usepackage{afterpage}
\usepackage{gensymb}
\usepackage{cancel}
\usepackage{color,bm}
\usepackage{longtable}
\usepackage[titletoc]{appendix}
\usepackage[nottoc]{tocbibind}
\usepackage{cases}
\usepackage{caption}
\usepackage{enumerate}
\usepackage{changes}
\usepackage{graphicx}
\usepackage{subfigure}
\usepackage[square,sort,comma,numbers,super]{natbib}
\usepackage{ulem}

\newcommand{\Eq}{Eq.~}
\newcommand{\Eqs}{Eqs.~}
\newcommand{\Fig}{Fig.~}
\newcommand{\Figs}{Figs.~}
\newcommand{\Tab}{Tab.~}

\newcommand{\jam}{{\mbox{\scriptsize J}}}

\newcommand{\inert}{{\mbox{\scriptsize inert}}}
\newcommand{\inte}{{\mbox{\scriptsize int}}}
\newcommand{\qs}{{\mbox{\scriptsize qs}}}
\newcommand{\hard}{{\mbox{\scriptsize hard}}}
\newcommand{\soft}{{\mbox{\scriptsize soft}}}
\newcommand{\rigi}{{\mbox{\scriptsize rig}}}
\newcommand{\defo}{{\mbox{\scriptsize def}}}
\newcommand{\el}{{\mbox{\scriptsize el}}}
\newcommand{\fluid}{f}
\newcommand{\solid}{s}

\begin{document}

\begin{center}
\LARGE{Merging fluid and solid granular behavior}\\[0.4cm]

\large{Dalila Vescovi\textit{$^{\ddag}$} and Stefan Luding}\\[0.4cm]

\normalsize{\textit{~Multi Scale Mechanics (MSM), CTW, MESA+, University of Twente, PO Box 217, 7500 AE Enschede, the Netherlands. \\
E-mail: d.vescovi@utwente.nl; s.luding@utwente.nl}}\\[8pt]
\normalsize{\ddag~Present address: Department of Civil and Environmental Engineering, Politecnico di Milano, 20133 Milano, Italy. E-mail: dalila.vescovi@polimi.it}
\end{center}

\vskip 1 cm

\begin{abstract}
Simple homogeneous shear flows of frictionless, deformable particles are studied by particle simulations at large shear rates and for differently soft, deformable particles.
The particle stiffness sets a time-scale that can be used to scale the physical quantities;
thus the dimensionless shear rate, i.e. the inertial number $I$ (inversely proportional to pressure), can alternatively be expressed as inversely proportional to the square root of the particle stiffness.
Asymptotic scaling relations for the field variables pressure, shear stress and granular temperature are inferred from simulations 
in both fluid and solid regimes, corresponding to unjammed and jammed conditions. Then the limit cases are merged to unique
constitutive relations that cover also the transition zone in proximity of jamming.
By exploiting the diverging behavior of the scaling laws at the jamming density, we arrive at continuous and differentiable phenomenological constitutive relations for the stresses and the granular temperature as functions of the volume fraction, shear rate, particle stiffness and distance from jamming. 
In contrast to steady shear flows of hard particles the (shear) stress ratio $\mu$ does not collapse as a function of the inertial number,
indicating the need for an additional control parameter. In the range of particle stiffnesses investigated, in the solid regime, 
only the pressure is rate independent, whereas the shear stress exhibits a slight shear rate- and stiffness-dependency.
\end{abstract}

%

\section{Introduction}\label{Intro}

In the recent past, the flow of granular materials has been the subject of many scientific works; this is due to the large number of natural phenomena (i.e., landslides and debris flows) and industrial processes involving solid particles flowing.
Granular materials can behave very differently depending on both the micro-mechanical properties of the particles and the macroscopic characteristic of the system (i.e., geometry, gradients of velocity, density). 
In particular, granular systems, as well as other soft matter such as suspensions of soft particles or polymers, foams, or emulsions, can behave like fluids, meaning that they yield under shear stress, or like solids able to resist applied stresses without deforming.\cite{sil2010}
Granular materials exhibit solid-like behavior if the particles are packed densely enough and a network of persistent contacts develops within the medium, resulting in a jammed mechanically stable structure of the particles.
On the other hand, when the grains are widely spaced and free to move in any direction, interacting only through collisions, the medium is unjammed and behaves like a fluid. \\
As a consequence, a key question concerns how to model the transition from fluid- to solid-like behavior and vice-versa? This transition is reflected in the relation between stresses and deformation rates, that is the rheology. 
Jammed structures can show a rate independent behavior, but in unjammed granular systems the stresses are proportional to the square of the strain rate (Bagnold scaling\cite{bag1954}). In the proximity of jamming, a continuous transition between the two extreme regimes takes place, for which the correct rheology is still not fully understood.\cite{lud2016}\\
Numerical simulations performed by using perfectly rigid spheres (i.e., infinite stiffness)\cite{mit2007,chi2013} have shown that pressure, shear stress and granular temperature (defined as the mean squared velocity fluctuations, representing a measure of the degree of agitation of the system) diverge when the density approaches jamming, as predicted by the kinetic theory of granular gases.\cite{gar1999,ves2014,ber2015} As a consequence, jammed flows of rigid particles are not possible at constant volume.
In contrast, numerical results with soft particles\cite{cam2002,ji2006,chi2012} show that steady, constant volume flows are possible also at densities above jamming. Their softness allows the particles to deform and to attain steady shear at very dense, jammed configurations.
Experiments performed on concentrated suspension \cite{van2013} of hard and rigid particles have revealed different physical processes in the straining and the flow behavior of hard and soft spheres suspension, in particular in the unsteady regime, due to the development of permanent contacts of the particles in the case of deformable spheres at large volume fractions.\\
Although several constitutive models have been proposed in the literature to account for the softness of the particles,\cite{chi2012,ber2011b,lud2009,nes2015,ber2015b} most of them have some important limitations. In particular, most of them ``match'' the limits of fluid- and solid-like behavior and many do not provide continuous and differentiable equations for all the variables of the system (i.e., pressure, shear stress and granular temperature). Many models are not able to quantitatively predict steady shear flows at all densities. \\
For a collection of particles, the jamming transition occurs in the limit of zero confining pressure at the critical volume fraction $\nu_\jam$ (where, for a granular material composed of identical particles, the local volume fraction $\nu$ is defined as the ratio of the local material density to the particle density, and the subscript $\jam$ refers to the jamming transition). 
Recently, several authors have shown that, under homogeneous steady shearing, stresses and granular temperature can be expressed as asymptotic power-law relations of the shear strain, if scaled by powers of $\left|\nu - \nu_\jam\right|$, the distance to jamming.\cite{ols2007,hat2008,ots2009,ots2010} However, different authors report different critical exponents in the solid (jammed) and fluid (unjammed) regimes, both numerically\cite{ols2007,hat2008,chi2012} and theoretically.\cite{ots2009} The critical exponents, as well as the jamming volume fraction, are affected by several parameters: the polydispersity of the system, \cite{oga2013,kum2014} the friction of the particles, the force contact model adopted in the simulations (linear spring or Hertzian) and the ratios of relevant time-scales like the inverse shear rate or square root of the particle stiffness per mass. 
Even though the jamming density is known to be pressure-, and material- as well as protocol-dependent, see Ref.~\cite{lud2016} and references therein, for the purpose of scaling we have to work with a constant $\nu_\jam$.
Jamming transition and scaling of quantities with respect to the distance from jamming are also very popular subjects in the fluid-dynamics and statistical-mechanics community. In particular, complex fluids composed of a dispersion of one material in a continuous phase show a transition from mechanically solid-like to fluid-like states, similarly to granular systems, when the shear stress is increased above some critical value, the yield stress.\cite{par2013} Emulsions, colloids, foams, gels and suspensions of (soft) particles or polymers belong to this category. 
By performing experiments on different kinds of yield-stress fluids, \citet{par2013} and \citet{din2015} have found that, by appropriate scaling with the distance to jamming, such kind of complex fluid systems allowed a data collapse onto universal power-law relations of the shear strain, in the fluid-like and solid-like regimes. The presence of the ``continuum phase'', i.e. the liquid in which droplets or particles are dispersed, leads to critical exponents for the case of complex fluids which differ from those of dry collections of soft particles, especially in the fluid-like state, but the intrinsic rheology is actually surprisingly analogous.\\
This work focuses on the simple shear flow of an ideal granular material, composed of identical, frictionless spheres, under steady conditions, at constant volume. 
A series of Discrete Element Method simulations are performed in order to investigate the role of particle stiffness in a wide range of volume fractions, both well below and above jamming. The goal is to propose phenomenological constitutive relations for granular materials merging dilute and dense flow conditions.
First, we analyze our numerical results to derive appropriate scaling laws in both unjammed and jammed states, far from the jamming transition (with special functional forms that are needed further on).
Then, we use the scaling laws that collapse the data in either solid or fluid regimes to postulate a phenomenological model for the stresses and the granular temperature. In particular, the two regimes are merged in a unique function which is (i) continuous and differentiable at any point and (ii) able to predict the behavior even close to the jamming transition. \\
This paper is organized as follows: in Section 2 we describe the simulation method and the flow configuration; in Section 3 the difference between rigid and soft particles granular systems are discussed in terms of the $\mu-I$ rheology\cite{gdr2004} and the jamming density is evaluated based on the behavior of the coordination number as function of volume fraction and particles stiffness. Section 4 is devoted to the scaling relations obtained from the numerical data; in Section 5 the merged constitutive model is presented and compared with numerical results; in Section 6 we discuss the comparison of the proposed model with the rheologies proposed by \citet{chi2012}, \citet{ber2015b}, \citet{sin2015} and \citet{par2013}; in particular, the model in Ref.~\cite{par2013} was proposed for emulsion-type systems, and is here revised to adapt to the case of dry assemblies of soft spheres, according to our numerical data. Finally, the results are summarized and concluding remarks are given in Section 7.

\section{DEM numerical simulations}

%
\begin{figure}[!h]
\centering
\includegraphics[width=4cm]{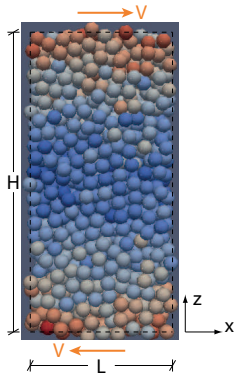}
\caption{Sketch of the constant-volume simple shear flow configuration. A granular material composed of frictionless, soft particles is homogeneously sheared at constant shear rate $\dot\gamma$, with flow in horizontal direction (walls move in opposite horizontal directions with velocity $V = \dot\gamma H/2$), by using Lees Edwards boundary conditions. Colors indicate speed, from blue (zero velocity of the particles in the core of the domain) to red (maximum velocity of the particles at the boundaries).}
\label{simpleshear_fig}
\end{figure}

\noindent Simple shear flows are homogeneous if the horizontal velocity of the medium is linearly changing along a line perpendicular to the shearing direction and the kinematic variable which affects the problem is its first spatial derivative, the shear rate, $\dot\gamma$, that is then constant along the flow depth.
The other variables governing the problem are the volume fraction $\nu$ (or density, or concentration, defined as the fraction of volume occupied by the spheres), the pressure $p$, the shear stress $s$ and the granular temperature $T$. The latter is defined as one third of the mean squared particle velocity fluctuations and represents a measure of the degree of agitation of the system, as introduced in the framework of kinetic theories of granular gases. \cite{jen1983,lun1991,cam1990,gar1999,gol2003}
\\
\noindent We have performed DEM numerical simulations of steady simple shear flow of frictionless spheres to investigate the role of the particle stiffness $k$, especially at large volume fraction.
The simulations and the coarse-graining presented in this paper, were undertaken using the open-source code Mercury-DPM (\href{www.mercurydpm.org}{www.mercurydpm.org}).\cite{tho2012,wei2012} 
The simulations are done under constant volume with a uniform shear in a rectangular box of dimensions $L\times L \times H$, respectively in $x$, $y$ and $z$ directions (\Fig\ref{simpleshear_fig}).
The shear is applied using Lees-Edwards\cite{lee1972} periodic boundary conditions in the $z$ direction and periodic boundary conditions are employed in the $x$ and $y$ directions.
In all simulations, we use 2000 particles of diameter $d$, density $\rho_p$, i.e., mass $m_p = \rho_p \pi d^3 /6$, and fix the height of the computational domain as $H = 20d$, before we compute the $x-$ and $y-$ size $L$ accordingly to the chosen volume fraction $\nu$. 
The system is sheared applying a constant velocity gradient (shear rate) $\dot\gamma = \partial u_x/\partial z$ with mean flow $u_x$ in the $x$ direction only. 
We use a linear spring-dashpot model, then the normal force between particles at contact is computed as $f_n = k_n \delta + \gamma_n \dot\delta$, with overlap $\delta$.\cite{lud1998,lud2008} $\gamma_n$ is the damping coefficient and is related to the normal coefficient of restitution: $\gamma_n = \sqrt{\left[4 m_p k_n \left(\log e_n\right)^2\right]/\left[\pi^2 + \left(\log e_n\right)^2\right]}$.
In the simulations, we use normal coefficient of restitution $e_n = 0.7$, tangential coefficient of restitution $e_t = -1$, interparticle friction coefficient $\mu = 0$ and normal spring stiffness $k_n = k$. The collision time can be analytically obtained as $\sqrt{\rho_p\pi d^3\left[\pi^2+\left(\log e_n\right)^2\right]/(12 k)}$.
Simulations have been performed by systematically changing both the volume fraction $\nu$, ranging from 0.2 to 0.68, and the particle stiffness $k$, such that the dimensionless quantity $k/\left(\rho_p d^3 \dot\gamma^2\right)$ ranges from $10^3$ to $10^7$. \\
At large volume fraction (i.e., $\nu > 0.64$) and large particle stiffness (i.e., $k/\left(\rho_p d^3 \dot\gamma^2\right) \geq 10^5$), we observed crystallization of the monodisperse systems, which results in non-homogeneous flows such as shear bands. To avoid crystalline structures, in such situations we use slightly polydisperse systems, with mean particle diameter $d$ 
 and uniform polydispersity $w = 1.2$, where $w$ is the ratio of the maximum to the minimum particle diameter. Such a small value of the polydispersity does not affect significantly the measured quantities, and, in particular, does not vary so much the critical volume fraction.\cite{kum2014} \\
Here and in the following, we use standard notation to refer to dimensional variables, whereas the star is adopted to denote quantities scaled using the particle diameter $d$, density $\rho_p$ and stiffness $k$. Then, the scaled granular temperature, pressure, shear stress and shear rate are given, respectively, as: $T^* = T d \rho_p / k$, $p^* = p d/k$, $s^* = s d/k$ and $\dot\gamma^* = \dot\gamma \sqrt{\rho_p d^3/k}$.
We also introduce three time scales: 
the microscopic relaxation time scale associated with the transversal motion of a particle submitted to a pressure $p$: $t_m = d\sqrt{\rho_p/p}$; 
the macroscopic time scale associated with the shear rate parallel to the flow: $t_\gamma = 1/\dot\gamma$; 
and the time scale associated with the particles deformability (contact duration): $t_c =\sqrt{\rho_p d^3/k}$.

\section{Influence of the particle stiffness}

\begin{figure}[!h]
\centering
\includegraphics[width=1.\textwidth]{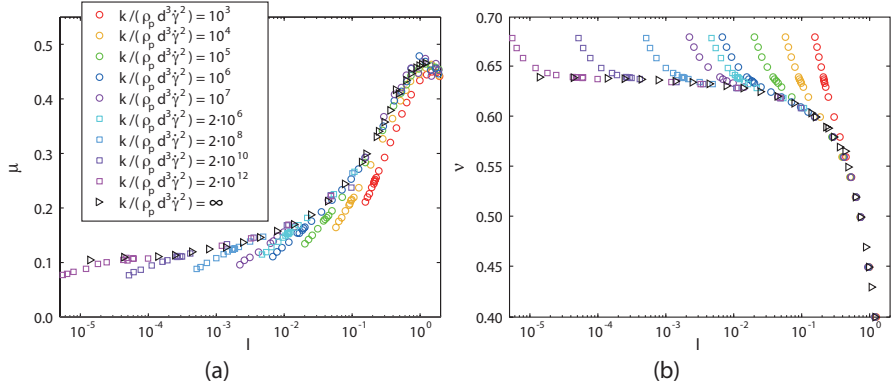}
\caption{Stress ratio (a) and volume fraction (b) versus the inertial number, for different values of the dimensionless particle stiffness. Circles, squares and triangles represent, respectively, our simulations, the numerical measurements of \citet{ots2009} and the data for rigid particles obtained by \citet{mit2007} and \citet{pey2008}.}
\label{muIrheo}
\end{figure}

A convincing yet simple phenomenological model, which has been used frequently in the literature in the last years, is the $\mu-I$ rheology.\cite{gdr2004,dac2005,jop2006} According to this model, only three dimensionless variables are relevant for steady shear flows of granular materials: the volume fraction $\nu$, the stress ratio $\mu = s/p$ and the inertial number $I$.
The inertial number represents the ratio between the microscopic relaxation and the macroscopic shear time scales: $I = t_m /t_\gamma = \dot\gamma d \sqrt{\rho_p/p}$\footnote[2]{Note that some authors define the inertial number using the bulk density $\rho = \rho_p \nu$ instead of the particle density: $I_b = \dot\gamma d \sqrt{ \rho / p } = \dot\gamma d \sqrt{ \rho_p \nu / p }  = I \sqrt{ \nu }$.}. 
On the planes $\nu-I$ and $\mu-I$, data obtained on different flow configurations collapse in the limit of rigid particles.\cite{for2008} Two main assumptions at the basis of the $\mu-I$ rheology are: (i) perfectly rigid particles and (ii) homogeneous flow (local rheology). Under these two assumption, $t_m$ and $t_\gamma$ are the only time scales influencing the problem.
When the particles are soft (`softness' effect\cite{chi2012}) and/or the flow is not homogeneous (`non-local' effects\cite{kam2012}), the standard $\mu-I$  laws fail because different mechanisms come into play.\\ 
In the case of non-homogeneous flows, boundaries affect the system and flow gradients cannot be neglected anymore. In such situations, \citet{kam2012} have proposed a new ``diffusive'' state parameter, the granular fluidity, with a diffusive evolution equation, to account for some non-locality (see also Ref.~\cite{goy2008,dec2015}).\\
On the other hand, when the particles are not perfectly rigid, contacts are not instantaneous but take a finite time $t_c$ during which a part of the energy, the elastic potential energy fraction, is stored\cite{lud2009} due to the persistent deformations of the particles. \\
When a steady flow is unjammed and sheared very slowly, rigid and deformable particles exhibit very similar behavior, controlled only by $t_m$ and $t_\gamma$, because $t_c$ is much smaller than either; when the transition to the solid, jammed phase takes place (and this happens only to soft spheres), the time scale associate to the particles deformability $t_c$ (contact duration) starts to influence the system, and softness effects appear. 
In this work, we investigate homogeneous steady shear flow of soft particles, but still can disregard non-local effects. \\
In \Fig\ref{muIrheo} we plot the stress ratio (a) and the volume fraction (b) versus the inertial number, measured from our simulations, for different values of the dimensionless particle stiffness $k/\left(\rho_p d^3 \dot\gamma^2\right)$, that corresponds to $\left(t_\gamma / t_c\right)^2$ (or, equivalently, $\dot\gamma^{*-2}$). 
In particular, we use $k/\left(\rho_p d^3 \dot\gamma^2\right)$ ranging from $10^3$ to $10^7$ (circles). 
Also plotted in \Fig\ref{muIrheo} are the numerical results of \citet{ots2009}, who adopted more rigid particles having $k/\left(\rho_p d^3 \dot\gamma^2\right)$ from $2\cdot 10^6$ to $2\cdot 10^{12}$ (squares, here we consider their data obtained using 20000 monodisperse spheres), and those of \citet{mit2007} and \citet{pey2008}, which were performed using rigid particles (triangles). Note that in our simulations and in Ref.~\cite{mit2007} the coefficient of normal restitution is 0.7, whereas in Ref.~\cite{ots2009} and Ref.~\cite{pey2008} it is very close to zero. This, however, does not affect the main features of the plots on the $\mu-I$ and $\nu-I$ planes, as discussed next.\\

\noindent As predicted by the $\mu-I$ rheology, in the case of infinitely hard particles, for vanishing $I$, the stress ratio tends to a constant, asymptotic value, often interpreted as the yield stress ratio. At the same time, the volume fraction saturates to the rigid-limit jamming value (here $\nu_\jam = 0.634$ as detailed in the following). 
In the case of deformable particles, the data deviate from the rigid limit curves and behave very differently depending on the softness. Data with large stiffness agree with the rigid limit data in a large range of inertial numbers and deviate for very slow shear conditions (i.e., very low $I$ due to either small $\dot\gamma$ and/or large $p$). 
For decreasing $I$ and constant stiffness, on the $\mu-I$ plane, \Fig\ref{muIrheo}(a), curves do not saturate but continuously decrease; the smaller the stiffness, the larger the $I$ when the softness effect arises. The transition to solid overcompressed jammed states ($\nu > \nu_\jam$ in \Fig\ref{muIrheo}b) is connected to this decrease in the stress ratio occurring for soft spheres. 
Note that the deviation from the standard $\mu-I$ rigid-particle rheology cannot be due to non-local effects, since all simulations presented are carried out in uniform conditions, i.e. non-homogeneous flow data are disregarded here.
Steady flows of very soft particles are possible only at large inertial numbers: for example, for $k/\left(\rho_p d^3 \dot\gamma^2\right) = 10^5$, $I$ never reaches $10^{-2}$.\\
In Ref.~\cite{sin2015} and Ref.~\cite{roy2016}, the authors observed a similar non-collapse of the stress ratio with the inertial number. They performed steady state flow simulations using soft (slightly frictional) particles in a split-bottom geometry, in which  gradients in stresses arise due to gradients in both strain rate and pressure. As a consequence, both non-locality and softness can affect the system and distinguishing between the two effects is not possible. 
In Ref.~\cite{chi2012} homogeneous simple shear flows of deformable, frictional particles have been carried out, and the deviation of the data from the standard $\mu-I$ rheology was reported only at large inertial numbers, $I \geqslant 10^{-1}$. Conversely, for frictionless spheres, we have observed deviations in the whole range of inertial numbers investigated. \\

\begin{figure}[!h]
\centering
\includegraphics[width=0.5\textwidth]{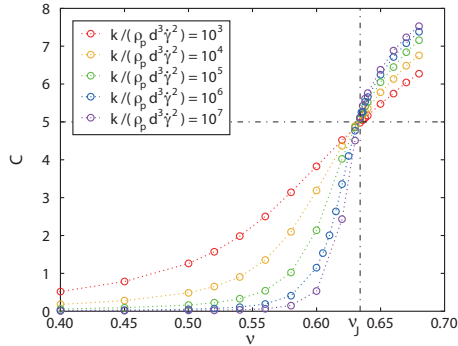}
\caption{Coordination number versus volume fraction, for different values of the dimensionless particle stiffness.}
\label{CoordNumber}
\end{figure}

\noindent \Fig\ref{CoordNumber} shows the mean coordination number $C$ (average number of contacts between all particles) in our simulations plotted versus the volume fraction. As observed for frictional spheres,\cite{ji2006} at low volume fraction, the coordination number increases with decreasing  particle stiffness. Decreasing stiffness is equivalent to an increasing scaled shear rate $\dot\gamma^*$. This dependency inverts at higher volume fraction.
Intuitively, when the system is unjammed the softness makes the contact duration longer resulting in a larger $C$, at a fixed density. On the other hand, when the system is jammed very deformable particles (small stiffness) allow to reach denser configurations than more rigid particles for a fixed value of the average number of contacts.\\
All the curves intersect each other at $\nu \cong 0.634$, where $C\approx 5$.
In theory, the jamming transition occurs at the isostatic point\cite{ohe2002,ohe2003,sil2002} which corresponds to a coordination number equal to 6.
In non-sheared isostatic packings, particles that do not belong to the force network (i.e. with exactly zero contacts) do not contribute to the coordination number. 
In shearing conditions, there may be particles having a finite number of contacts for some short time, which do not contribute to the mechanical stability of the packing.
Frictionless particles with less than 4 contacts are considered ``rattlers'', since they cannot be mechanically stable and hence do not contribute to the contact network.\cite{gon2009,kum2014} The ``corrected'' coordination number $C^*$ has been introduced as the ratio between the total number of contacts of the $N_4$ particles with at least 4 contacts, and the rattler fraction is defined as $\phi_r = (N-N_4)/N$, where $N$ is the total number of particles in the system.
In Ref.~\cite{gon2010}, the authors have shown that the corrected and the standard coordination number are related as $C = C^*(1-\phi_r)$. In particular, at jamming the corrected coordination number is 6, as in isostatic conditions, and $\phi_r = 0.13 \pm 0.03$, then the standard coordination number results $C = 5.22 \pm 0.18$.
This value of $C$ is consistent with that at which all the curves intersect and change convexity in \Fig\ref{CoordNumber}, denoting the transition between solid, jammed structures and fluid, unjammed systems. As a consequence, we assume the corresponding value of $\nu$ to be the jamming volume fraction: $\nu\jam = 0.634$.\\
In general, the jamming volume fraction is not a constant (even if pressure or density are held constant),\cite{kum2016,lud2016} but shows a history-, material-, and protocol-dependence. We cannot exclude \emph{a priori} that also particle stiffness and shear rate affect the jamming density. In our simulations, variability of $\nu_\jam$ is small relative to the full range of data investigated, but, in general, it could make a difference, especially for soft particles. Nevertheless, in this work we assume, for simplicity, the jamming volume fraction to be constant and equal to 0.634.

\section{Scaling plots}

As previously stated, two extreme flow behaviors, commonly interpreted as flow regimes, are associated with different power laws scaling with $\dot\gamma^*$: 
stresses scale as $\dot\gamma^{*2}$ in the fluid, unjammed regime, that is below the critical volume fraction ($\nu < \nu_\jam$), but show almost no rate dependence in the solid, jammed regime ($\nu > \nu_\jam$). At $\nu \sim \nu_\jam$, a continuous transition between the two extreme regimes takes place (jamming transition).\\
\noindent In order to find out the scaling functions for the stresses and the temperature, we introduce the following scaling relation for the generic scaled quantity $X^*$ (with $X^*$ being a place-holder for either $p^*$, $s^*$ or $T^*$):
\begin{equation}\label{X_gp}
\dfrac{X^*\nu^{\lambda}}{\left|\nu-\nu_\jam\right|^{\alpha}} \sim \left[ \dfrac{\dot\gamma^*}{\left|\nu-\nu_\jam\right|^{\beta}} \right]^{m^i},
\end{equation}
where $\alpha$, $\beta$ and $\lambda$ are constant coefficients, whereas the exponent $m^i$ depends on the flow regime $i$. 
In this framework, we do not consider the scaling relation to describe the jamming transition. 
Assuming $i = \solid$ in the solid and $i = \fluid$ in the fluid regime; far from the jamming volume fraction, \Eq\eqref{X_gp} can be expressed as
\begin{equation}\label{X_gp2}
X^*\nu^{\lambda} = \begin{cases}
x_\fluid\dfrac{\dot\gamma^{*m^{\fluid}}}{\left(\nu_\jam-\nu\right)^{q^{\fluid}}}, & \mbox{if } \nu < \nu_\jam\\
x_\solid\dot\gamma^{*m^{\solid}}\left(\nu-\nu_\jam\right)^{q^{\solid}}, & \mbox{if } \nu > \nu_\jam,
\end{cases}
\end{equation}
where $q^{\fluid} = \left(m^{\fluid}\beta-\alpha\right)$, $q^{\solid} = \left(\alpha-m^{\solid}\beta\right)$ and $x_\fluid$ and $x_\solid$ are dimensionless constitutive parameters, such that the $\dot\gamma^*$ scaling is encompassed by $m^i$ and the density scaling by $q^i$.\\

\noindent Next, we determine the critical exponents and the constant coefficients for the granular temperature and the stresses by using the results obtained from our numerical simulations with frictionless spheres. The constant coefficients are very sensitive to both the critical volume fraction $\nu_\jam$ and the range of scaled shear rate $\dot\gamma^{*}$ analyzed.\cite{ots2009,chi2012} 
In our simulations, the scaled shear rate ranges from $10^{-4}$ to $10^{-2}$ ($\dot\gamma^{*}= \left[k/\left(\rho_p d^3 \dot\gamma^2\right)\right]^{-1/2}$) and, for all the quantities, we find good collapses  using $\nu_\jam = 0.634$, the jamming volume fraction obtained in the previous Section. The extrapolated values of the coefficients for the stresses and the granular temperature resulting from our DEM results are reported in \Tab\ref{tab1}. 
The values in \Tab\ref{tab1} are estimated by fitting our numerical data, and are quite similar to those obtained in Ref.~\cite{hat2008} for bidisperse mixture of frictionless particles.\\

\begin{table}[h]
\begin{center}
\caption{\ Scaling exponents appearing in \Eq\eqref{X_gp} for pressure, shear stress and granular temperature, as inferred from the numerical simulations}
\label{tab1}
\begin{tabular}{cccccccc}
\hline
$X^*$ & $\alpha$ & $\beta$ & $\lambda$ & $m^\fluid$ & $m^\solid$ & $q^{\fluid}$ & $q^\solid$  \\
\hline
$p^*$ & 6/5 & 9/5 & 1 & 2 & 0 & 12/5 & 6/5  \\
$s^*$ & 6/5 & 8/5 & 1/2 & 2 & 1/6 & 2 & 14/15  \\
$T^*$ & 2 & 3/2 & 2 & 2 & 1 & 1 & 1/2  \\
\hline
\end{tabular}
\end{center}
\end{table}

\noindent Then, the resulting relations for the scaled pressure, shear stress and granular temperature become, respectively:
\begin{align}
\dfrac{p^*\nu}{\left|\nu-\nu_\jam\right|^{6/5}} \sim \left[ \dfrac{\dot\gamma^*}{\left|\nu-\nu_\jam\right|^{9/5}} \right]^{m_p^i}, \label{p_gp}\\
\dfrac{s^*\nu^{1/2}}{\left|\nu-\nu_\jam\right|^{6/5}} \sim \left[\dfrac{\dot\gamma^*}{\left|\nu-\nu_\jam\right|^{8/5}} \right]^{m_s^i}, \label{s_gp}\\
\dfrac{T^*\nu^2}{\left|\nu-\nu_\jam\right|^2} \sim \left[\dfrac{\dot\gamma^*}{\left|\nu-\nu_\jam\right|^{3/2}}\right]^{m_T^i}. \label{T_gp}
\end{align}

\begin{figure}[!h]
\centering
\includegraphics[width=1.\textwidth]{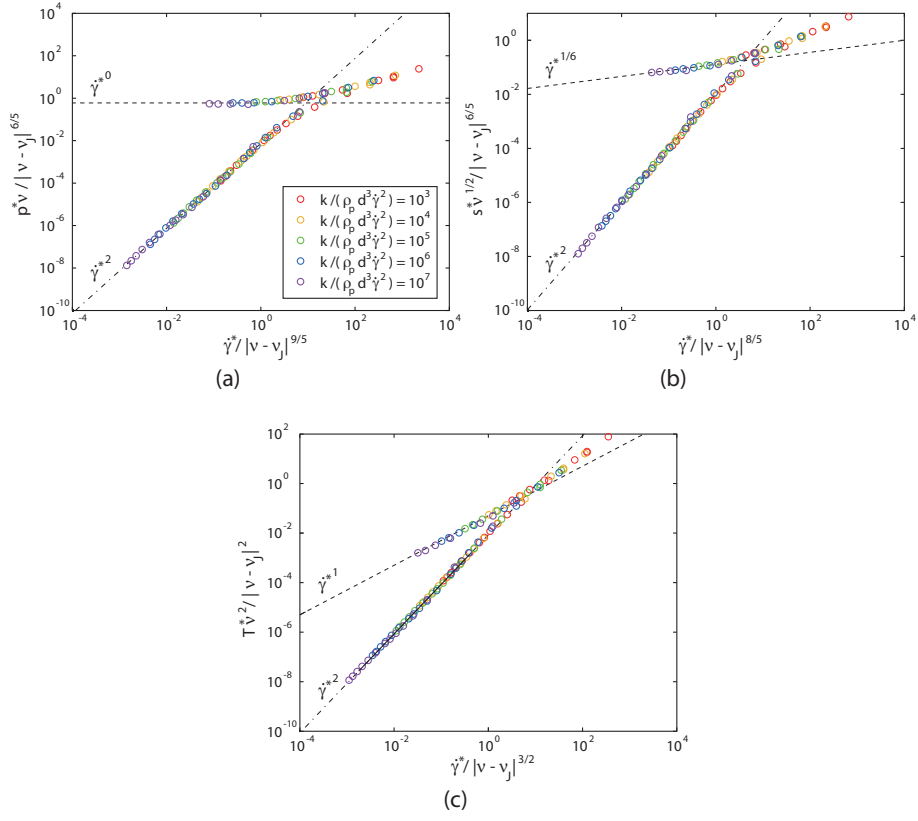}
\caption{Collapse of scaled (a) pressure, (b) shear stress and (c) granular temperature plotted against scaled shear rate, for different values of the (dimensionless) particle stiffness as given in the inset. Dash-dotted lines represent the scaling laws in the unjammed regime whereas dashed lines those in the jammed regime. The right most data are those for $\nu \sim \nu_\jam$ and are not supposed to collapse on the limit cases.}
\label{scaling_fig}
\end{figure}

\noindent \Figs\ref{scaling_fig}(a)-(c) show, for the three quantities, the collapse of numerical data obtained using \Eqs\eqref{p_gp}-\eqref{T_gp}, respectively, using $\nu_\jam = 0.634$. 
From \Figs\ref{scaling_fig}(a)-(c), we obtain the values of the critical exponents $m_p^i$, $m_s^i$ and $m_T^i$ in both the fluid (unjammed) and the solid (jammed) regime, by interpolating the  collapsing curves at small enough $\dot\gamma^*$. 
In particular,
\begin{itemize}
\item when $\nu < \nu_\jam$ (fluid regime)\\
\[ m_p^{\fluid} = m_s^{\fluid} = m_T^{\fluid} = 2; \]
\item when $\nu > \nu_\jam$ (solid regime)\\
\[ m_p^{\solid} = 0, \: m_s^{\solid} = 1/6, \: m_T^{\solid} = 1. \]
\end{itemize}

\begin{table}[!h]
\begin{center}
\caption{\ Constant coefficients appearing in \Eqs\eqref{lim_pstar}-\eqref{lim_Tstar} as inferred from the numerical simulations}
\label{tab1_2}
\begin{tabular}{ccccccccc}
\hline
$\nu_\jam$ & $p_\fluid$ & $p_\solid$  & $s_\fluid$ & $s_\solid$ & $t_\fluid$ & $t_\solid$ & $\mu_\fluid$ & $\mu_\solid$ \\
\hline
0.634 & 0.0075 & 0.60  & 0.0105 & 0.12 & 0.0090 & 0.05 & 1.4 & 0.2 \\
\hline
\end{tabular}
\end{center}
\end{table}

\begin{figure}[!h]
\centering
\includegraphics[width=0.5\textwidth]{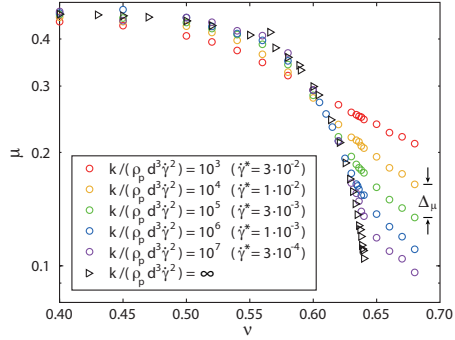}
\caption{Stress ratio as a function of the volume fraction for different values of the particle stiffness. Also plotted (black triangles) are the rigid particle data obtained by \citet{mit2007} and \citet{pey2008}.}
\label{munu_dat}
\end{figure}

\noindent As a consequence, for small $\dot\gamma^*$, we obtain:
\begin{align}
p^* &= \begin{cases}
	\dfrac{p_\fluid}{\nu}\dfrac{\dot\gamma^{*2}}{\left(\nu_\jam-\nu\right)^{12/5}}, & \mbox{if } \nu < \nu_\jam,\\
	\dfrac{p_\solid}{\nu}\left(\nu-\nu_\jam\right)^{6/5}, & \mbox{if } \nu > \nu_\jam;
\end{cases}\label{lim_pstar}\\
s^* &= \begin{cases}
	\dfrac{s_\fluid}{\nu^{1/2}}\dfrac{\dot\gamma^{*2}}{\left(\nu_\jam-\nu\right)^{2}}, & \mbox{if } \nu < \nu_\jam,\\
	\dfrac{s_\solid}{\nu^{1/2}}\dot\gamma^{*1/6}\left(\nu-\nu_\jam\right)^{14/15}, & \mbox{if } \nu > \nu_\jam;
\end{cases}\label{lim_sstar}\\
T^* &= \begin{cases}
	\dfrac{t_\fluid}{\nu^{2}}\dfrac{\dot\gamma^{*2}}{\left(\nu_\jam-\nu\right)}, & \mbox{if } \nu < \nu_\jam,\\
	\dfrac{t_\solid}{\nu^{2}}\dot\gamma^{*}\left(\nu-\nu_\jam\right)^{1/2}, & \mbox{if } \nu > \nu_\jam.
\end{cases}\label{lim_Tstar}
\end{align}
The dimensionless constitutive parameters appearing in \Eqs\eqref{lim_pstar}-\eqref{lim_Tstar} can be inferred fitting the power laws in \Figs\ref{scaling_fig}(a)-(c) and are reported in \Tab\ref{tab1_2}.\\
We point out that: 
\begin{itemize}
\item[(i)] in both the extreme regimes, $p^*$, $s^*$ and $T^*$ depend not only on $\dot\gamma^*$ and $\left|\nu-\nu_\jam\right|$, but also on some power of the volume fraction ($\nu^{-1}$, $\nu^{-1/2}$ and $\nu^{-2}$, respectively).
\item[(ii)] In the fluid regime, all the three quantities scale quadratically with the scaled shear rate (Bagnold scaling\cite{bag1954}), but with different (negative) powers of $\left(\nu_\jam-\nu\right)$. 
\item[(iii)] Differently from what was shown in other works, in the jammed regime only the pressure is rate independent, whereas the shear stress appears to scale with $\dot\gamma^{*1/6}$ but only in a very limited range. The granular temperature scales linearly with $\dot\gamma^*$, as already suggested in Ref.~\cite{hat2008,ots2009}.
\end{itemize}
The slight rate-dependency of the shear stress in the solid regime becomes evident when plotting the stress ratio $\mu = s/p$, also known as macroscopic friction, as a function of the volume fraction. 
As shown in \Fig\ref{munu_dat}, at small volume fraction, $\mu$ is not much affected by the particle stiffness (i.e. by $\dot\gamma^*$): the numerical measurements of $\mu$ obtained using different particle stiffnesses collapse together with the rigid particle data obtained by \citet{mit2007} and \citet{pey2008}, deviating only for the most soft particles. This means that pressure and shear stress scale with the same power ($m^f = 2$) of $\dot\gamma^*$ in the unjammed regime. Therefore, the stress ratio in unjammed configurations of homogeneous steadily sheared systems depends only on the volume fraction.
For increasing volume fractions, the curves deviate more and more from the rigid particle limit, already below the jamming volume fraction, and remain well separated at $\nu > \nu_\jam$. Furthermore, at larger volume fractions, the separation factor is almost the same between all the curves, and equal to $\Delta_\mu = 10^{1/12}$, suggesting that $\mu \sim k/\left(\rho_p d^3\dot\gamma^2\right)^{-1/12} \sim \dot\gamma^{*1/6}$ in the solid regime. As a consequence, pressure and shear stress must have different asymptotic power-law relations with the scaled shear rate; when $p^*$ is rate independent, $s^*$ turns out to be affected by $\dot\gamma^*$, and thus by $k$, in the jammed regime.

\section{Model}

The goal of this Section is to propose phenomenological constitutive relations for $p^*$, $s^*$ and $T^*$ based on the asymptotical analysis discussed in the previous Section.
The idea is to merge the equations describing the two scaling regimes with a unique function which must be (i) continuous and differentiable at any point and (ii) able to predict the behaviour even at and around the jamming transition.\\
In the fluid and solid regimes, pressure, shear stress and granular temperature are given by 
\Eqs\eqref{lim_pstar}-\eqref{lim_Tstar}; these equations show the same form, as summarized in \Eq\eqref{X_gp2}.
Both branches of \Eq\eqref{X_gp2} can be solved explicitly for the distance to jamming:
\begin{align}
\nu - \nu_\jam & = - \left(\dfrac{x_\fluid\dot\gamma^{* m^{\fluid}}}{X^*\nu^{\lambda}}\right)^{1/q^{\fluid}}, & \mbox{if } \nu < \nu_\jam, \label{XX1}\\
\nu - \nu_\jam & = \left(\dfrac{X^*\nu^{\lambda}}{x_\solid\dot\gamma^{*m^{\solid}}}\right)^{1/q^{\solid}}, & \mbox{if } \nu > \nu_\jam. \label{XX2}
\end{align}
By defining the distance to jamming as the sum of the two right hand side contributions of \Eqs\eqref{XX1} and \eqref{XX2}, we obtain a unique continuous and differentiable equation, valid for any value of the volume fraction:
\begin{equation}\label{xy}
\nu - \nu_\jam = \left(\dfrac{X^*\nu^{\lambda}}{x_\solid\dot\gamma^{*m^{\solid}}}\right)^{1/q^{\solid}} - \left(\dfrac{x_\fluid\dot\gamma^{* m^{\fluid}}}{X^*\nu^{\lambda}}\right)^{1/q^{\fluid}}.
\end{equation}
\Eq\eqref{xy} does not include singularities, but cannot, in general, be solved explicitly for $X^*$.
We can derive implicit functions which relate pressure, shear stress and temperature to the volume fraction, through the scaled shear rate, in the same form as \Eq\eqref{xy}:
\begin{align}
\nu - \nu_\jam &= \left(\dfrac{p^*\nu}{p_\solid}\right)^{5/6} - \left(\dfrac{p_\fluid\dot\gamma^{*2}}{p^*\nu}\right)^{5/12}, \label{pstar}\\
\nu - \nu_\jam &= \left(\dfrac{s^*\nu^{1/2}}{s_\solid\dot\gamma^{*1/6}}\right)^{15/14} - \left(\dfrac{s_\fluid\dot\gamma^{*2}}{s^*\nu^{1/2}}\right)^{1/2}, \label{sstar}\\
\nu - \nu_\jam &= \left(\dfrac{T^*\nu^{2}}{t_\solid\dot\gamma^*}\right)^{2} - \left(\dfrac{t_\fluid\dot\gamma^{*2}}{T^*\nu^{2}}\right). \label{Tstar}
\end{align}
It is important to notice that the form of the merging function \Eq\eqref{xy} is general and can be used for any quantity for which the asymptotical behavior in the fluid and solid regime is known and which is given in the form of \Eq\eqref{X_gp2}. In particular, different values of the critical exponents of $\left|\nu-\nu_\jam\right|$ can be adopted for $p^*$, $s^*$ and $T^*$, as well as different power of the volume fraction.\\
Moreover, combining the asymptotical equations for $p^*$ and $s^*$, \Eqs\eqref{lim_pstar}-\eqref{lim_sstar}, we obtain the limit equations for the stress ratio $\mu = s/p = s^*/p^*$:
\begin{equation}
\mu = \begin{cases}
	\nu^{1/2}\mu_\fluid\left(\nu_\jam-\nu\right)^{2/5}, & \mbox{if } \nu < \nu_\jam,\\
	\nu^{1/2}\mu_\solid\dfrac{\dot\gamma^{*1/6}}{\left(\nu-\nu_\jam\right)^{4/15}}, & \mbox{if } \nu > \nu_\jam,
\end{cases}\label{lim_mu}\\
\end{equation}
which, following the same approach as above, results in an implicit, merging function for $\mu$:
\begin{equation}\label{mustar}
\nu - \nu_\jam = \left(\dfrac{\mu_\solid\nu^{1/2}\dot\gamma^{*1/6}}{\mu}\right)^{15/4} - \left(\dfrac{\mu}{\nu^{1/2}\mu_\fluid}\right)^{5/2}
\end{equation}
with parameters $\mu_\fluid = s_\fluid/p_\fluid=1.4$ and $\mu_\solid = s_\solid/p_\solid = 0.2$.\\

\begin{figure}[!h]
\centering
\includegraphics[width=1.\textwidth]{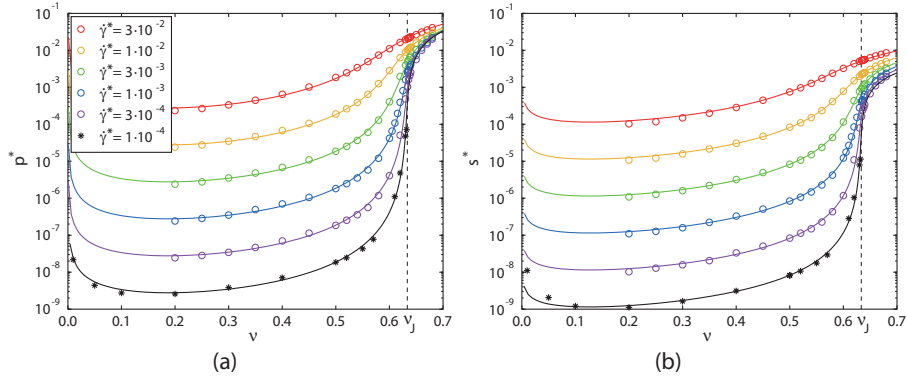}
\caption{Scaled (a) pressure and (b) shear stress as functions of the volume fraction for different values of the scaled shear rate $\dot\gamma^*$ (or, equivalently, stiffness: $\dot\gamma^* = t_c /t_\gamma \propto k^{-1/2}$). Symbols represent the numerical data obtained from numerical simulations and solid lines the proposed model \Eqs\eqref{pstar}-\eqref{sstar}. The black asterisks are data from Ref.~\cite{chi2013}.}
\label{ps_star_fig}
\end{figure}

\noindent In the following \Eqs\eqref{pstar}-\eqref{Tstar} and \eqref{mustar} are solved and compared with the numerical results of our simple shear simulations.\\
\Fig\ref{ps_star_fig} (a) and (b) depict the predicted scaled pressure and
shear stress, compared with the results of our numerical simulations of simple shearing for a range of scaled shear rates from $3\cdot 10^{-4}$ to $3\cdot 10^{-2}$. Also shown are the simple shear results of \citet{chi2013} (asterisks) for frictionless spheres having $e_n = 0.7$ and $\dot\gamma^* = 10^{-4}$.
The agreement is good for both variables in the whole range of volume fractions investigated and, in particular, in the transitional regime, i.e. around the jamming volume fraction $\nu_\jam = 0.634$. 
Only at very small volume fractions ($\nu < 0.1$) the shear stress is underestimated by the model.  
In the solid, jammed regime ($\nu > \nu_\jam$), the measured pressure data tend to collapse, in agreement with the rate independent behavior observed in other works,\cite{ji2008,hat2008,ots2009,chi2012} as predicted by \Eq\eqref{lim_pstar}, except for the softest particles ($\dot\gamma^* \geq 10^{-2}$). Conversely, when the contact duration gets into the order of magnitude of the shear time-scale, a slight systematic dependence of the shear stress on the shear rate is observed even at large volume fractions (\Fig\ref{ps_star_fig}b), which is well captured by the proposed model.

\begin{figure}[!h]
\centering
\includegraphics[width=1.\textwidth]{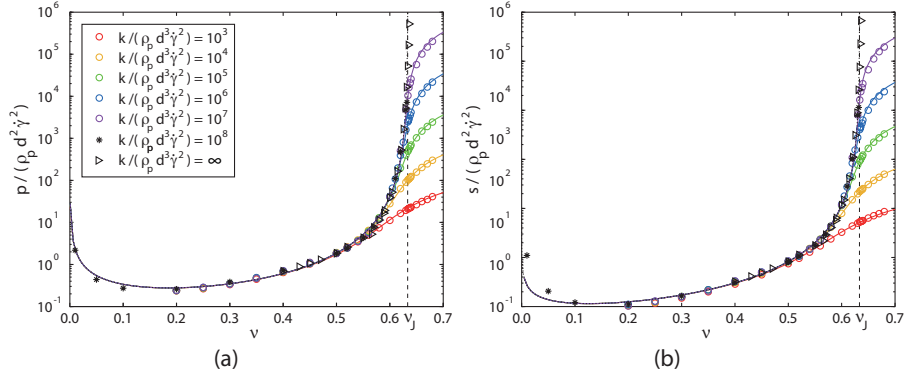}
\caption{Dimensionless (a) pressure $p/\left(\rho_p d^2 \dot\gamma^{2}\right) = p^*/\dot\gamma^{*2}$ and (b) shear stress $s/\left(\rho_p d^2 \dot\gamma^{2}\right) = s^*/\dot\gamma^{*2}$ as functions of the volume fraction for different dimensionless particle stiffness $k/\left(\rho_p d^3 \dot\gamma^{2}\right)$. Symbols represent data from the simulations and solid lines the proposed model \Eqs\eqref{pstar}-\eqref{sstar}. The black asterisks are data from Ref.~\cite{chi2013}; the black triangles are rigid data from Ref.~\cite{mit2007,pey2008} and the dotted lines are \Eqs\eqref{p_rig} and \eqref{s_rig}.}
\label{ps_fig}
\end{figure}

\noindent In \Fig\ref{ps_fig} (a) and (b), the (dimensional) pressure and the shear rate are made dimensionless using the particle diameter, density and the (dimensional) shear rate $\dot\gamma$. 
Different colors, as in \Fig\ref{ps_star_fig}, indicate different values of the dimensionless particle stiffness $k/\left(\rho_p d^3 \dot\gamma^{2}\right)$, ranging from $10^3$ to $10^7$. The results of the numerical simulations with rigid particles Ref.~\cite{mit2007,pey2008} are added to the figures, as well as the limit curves for rigid particles ($k\longrightarrow \infty$, i.e. $\dot\gamma^*\longrightarrow 0$, black dotted lines); these equations are identical to those of the unjammed regime in \Eqs\eqref{lim_pstar}-\eqref{lim_sstar} and diverge at the jamming volume fraction:
\begin{align}
\dfrac{p^\rigi}{\rho_p d^2\dot\gamma^2} & = \dfrac{p_\fluid}{\nu\left(\nu_\jam-\nu\right)^{12/5}},\label{p_rig}\\
\dfrac{s^\rigi}{\rho_p d^2\dot\gamma^2} & = \dfrac{s_\fluid}{\nu^{1/2}\left(\nu_\jam-\nu\right)^{2}}.\label{s_rig}
\end{align}
The predicted curves of $p/\left(\rho_p d^2\dot\gamma^2\right)$ and $s/\left(\rho_p d^2\dot\gamma^2\right)$ for soft particles collapse on the rigid ones in the fluid, unjammed regime, and start to deviate at different values of the volume fraction depending on the particle stiffness (even for the softest particles the deviation from the rigid limit is smaller than 15\% for $\nu < 0.55$, for both pressure and shear stress). The softer the particles, the lower the transition density.
As observed in \Fig\ref{ps_star_fig}(b), the shear stress is strongly underpredicted at very small densities, $\nu < 0.1$, where the factor $\nu^{-1/2}$ in \Eq\eqref{s_rig} strongly affects the expression of $s$, and is too small, not catching the functional behavior in the regime where kinetic theory holds.\cite{ves2013} However, the model allows for very good predictions for all the other values of volume fractions.

\begin{figure}[!h]
\centering
\includegraphics[width=1.\textwidth]{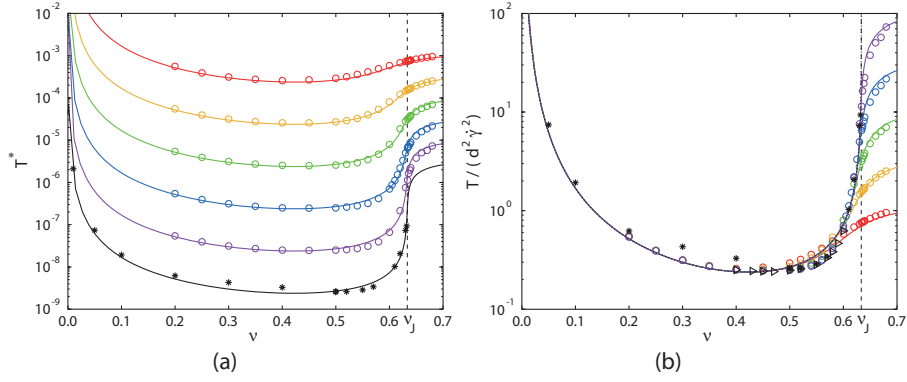}
\caption{Scaled and dimensionless granular temperature (a) $T^* = T\rho_p d/k$ and (b) $T/\left(d^2 \dot\gamma^{*2}\right) = T^*/\dot\gamma^{*2}$ as a function of the volume fraction for different values of the dimensionless particle stiffness or, equivalently, scaled shear rate [See legends in \Fig\ref{ps_star_fig} for (a) and in \Fig\ref{ps_fig} for (b)]. Symbols represent the data obtained from simulations and solid lines are the proposed model \Eq\eqref{Tstar}. The black asterisks and triangles are data from Ref.~\cite{chi2013} and Ref.~\cite{mit2007}, respectively; the dotted line is \Eq\eqref{T_rig}.}
\label{T_fig}
\end{figure}

\noindent The model predictions for the granular temperature are compared with the numerical results in \Fig\ref{T_fig} in terms of $T^* = T\rho_p d/k$ (a) and $T/\left(d^2\dot\gamma^2\right)$ (b). Like for the stresses, data with different stiffness collapse on the rigid curve in the fluid regime, at $\nu < 0.45$, if scaled with the particle diameter and the shear rate, see \Fig\ref{T_fig}(b).
According to \Eq\eqref{lim_Tstar}, the expression of $T$ in the rigid case is:
\begin{equation}\label{T_rig}
\dfrac{T^\rigi}{d^2\dot\gamma^2}  = \dfrac{t_\fluid}{\nu^2\left(\nu_\jam-\nu\right)}.
\end{equation}
On the other hand, \Fig\ref{T_fig}(a) highlights the linear rate dependence (inversely proportional to $k^{1/2}$) of $T^*$ in the solid regime, where the data do not collapse on a single curve (for large volume fractions).  
In \Fig\ref{T_fig}(b), for the softest particles the deviations from the rigid limit are smaller than $15\%$ for $\nu < 0.45$, whereas they are larger than $35\%$ for $0.45 \leq \nu \leq 0.55$. Hence, for the granular temperature, conspicuous softness effects arise already at volume fractions smaller than for pressure and shear stress, and are not properly reproduced by \Eq\eqref{Tstar}.\\

\begin{figure}[!h]
\centering
\includegraphics[width=1.\textwidth]{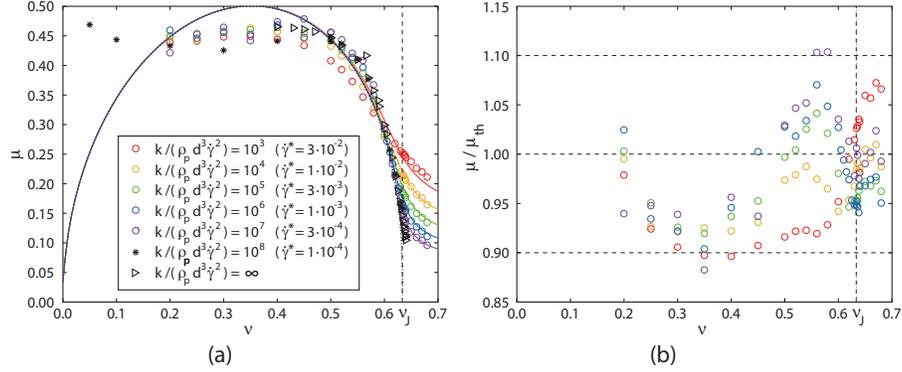}
\caption{(a) Numerical (symbols) and theoretical (lines, \Eq\eqref{mustar}) stress ratio plotted versus volume fraction, for different values of the dimensionless particle stiffness or, equivalently, scaled shear rate. (b) Numerical stress ratio $\mu$ scaled by the proposed prediction $\mu_{th}$, \Eq\eqref{mustar}. The disagreements are smaller than 10\% in the fluid range of volume fraction investigated, and are considerably better for the transitional and solid regime.}
\label{mu_fig}
\end{figure}

\noindent Finally, in \Fig\ref{mu_fig}, we compare the theoretical expression for the stress ratio \Eq\eqref{mustar} with the numerical measurements. The macroscopic friction $\mu$ varies over a narrow range so we can use a linear scale instead of a logaritmic one, differently from what was done for the other quantities, so that the quality of the predictions can be appreciated more accurately. 
From \Eq\eqref{mustar}, the stress ratio in the rigid limit reads:
\begin{equation}\label{mu_rig}
\mu^{\rigi}  = \mu_\fluid\nu^{1/2}\left(\nu_\jam-\nu\right)^{2/5}.
\end{equation}
In \Fig\ref{mu_fig}(a), the proposed model captures well the stress ratio, especially in the transition and solid regime. Some disagreements still remain in the fluid regime at volume fractions between 0.2 and 0.6, where the model does not capture well the slight $k$-dependence of $\mu$. Nevertheless, the disagreements are smaller than 10\% in the range of volume fractions investigated, see the quality factor in \Fig\ref{mu_fig}(b) where $\mu_{th}$ is the theoretical prediction of $\mu$ given by \Eq\eqref{mustar}. At $\nu < 0.2$, the model is unable to reproduce the increase of $\mu$ for decreasing $\nu$, and the stress ratio nullifies when $\nu = 0$, \Eq\eqref{mu_rig}, in contrast with the data. This discrepancy is due to the bad predictions in the stress ratio at very small volume fractions, and, in particular, due to the inappropriate exponent $-1/2$ of the volume fraction in \Eq\eqref{s_rig}.\\

\noindent As previously stated, the critical exponents strongly depend on the range of scaled shear rates. \citet{ots2009} performed homogeneous simple shear volume-controlled simulations of monodisperse, frictionless particles having stronger dissipation $e_n = 0.043$ and using a range of smaller $\dot\gamma^*$ from $7\cdot 10^{-7}$ to $7\cdot 10^{-4}$. They found exponents different from ours for the quantities, as well as a slightly different value of the jamming volume fraction (0.639 in contrast to our 0.634). Our numerical data thus do not scale with their theoretically predicted exponents, in neither fluid nor solid states. This discrepancy can be due to the different values of the coefficient of restitution adopted, which highlights the sensitivity of scaling results and exponents, in particular, to the magnitude of dissipation. 
Furthermore, the shear stress measured by Otsuki and Hayakawa exhibits a strictly rate-independent behavior in the solid state, in disagreement with our dependence on $\dot\gamma^*$, which, however, encompasses both rate and softness effects. 
We have previously discussed that a rate independent $s^*$ at $\nu > \nu_\jam$ implies that the stress ratio is only a function of the volume fraction, independent on $\dot\gamma^*$, in contrast with what is shown in \Fig\ref{munu_dat}.
On the other hand, if we try to predict their numerical results with our model, we do not succeed at very small $\dot\gamma^*$ (or, equivalently, very high stiffness). 
The choice to relate $s^*$, $\dot\gamma^*$ and $\left(\nu -\nu_\jam\right)$ through a power law in the solid, jammed state, while working for our intermediate regime of stiffnesses and shear rates, is probably not the right approach to reproduce flows in a wider range of shear rates, including their very slow and very stiff cases. 
Nevertheless, the form of the merging function proposed is suitable with any function of the shear rate, provided that it is proportional to some power of the distance to jamming. For example, if in the solid regime the scaled shear stress can be expressed as $s^* = g(\dot\gamma^*,\nu)\left(\nu-\nu_\jam\right)^q$, with $g$ an arbitrary function of $\nu$ and $\dot\gamma^*$, then the constitutive equation for $s^*$ simply reduces to $\nu - \nu_\jam = \left(\dfrac{s^*}{g(\dot\gamma^*,\nu)}\right)^{1/q} - \left(\dfrac{s_\fluid\dot\gamma^{*2}}{s^*\nu^{1/2}}\right)^{1/2}$. The lack of numerical data at larger stiffnesses (and small shear rates) in jammed conditions, due to very long computational time required, prevents us from finding a more appropriate relation for the shear rate in the solid regime, tending to a rate independent behavior in the limit of very large stiffness.\\

\begin{figure}[!h]
\centering
\includegraphics[width=1.\textwidth]{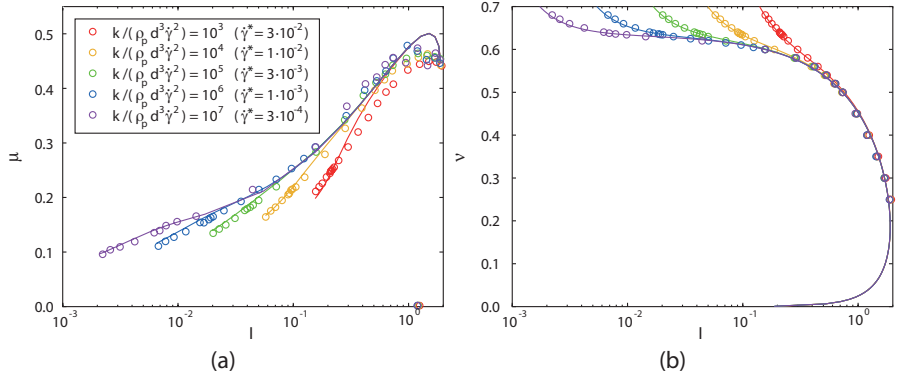}
\caption{Stress ratio (a) and volume fraction (b) versus the inertial number, for different values of the dimensionless particle stiffness. Symbols represent data from the simulations and solid lines the proposed model.}
\label{muI_figs}
\end{figure}

\noindent In \Fig\ref{muI_figs} we depict, for completeness, the comparisons between the model and the data on the planes $\mu-I$ (a) and $\nu-I$ (b), considering curves at constant $\dot\gamma^*$ (that is constant $k$). The volume fraction can be easily expressed as a function of $I$ and $\dot\gamma^*$ by substituting $I=\dot\gamma^*/p^{*1/2}$ into \Eq\eqref{pstar}:
\begin{equation}\label{nuIstar}
\nu -\nu_\jam = \left(\dfrac{\dot\gamma^{*2}\nu}{p_\solid I^2}\right)^{5/6} - \left(\dfrac{p_\fluid I}{\nu}\right)^{5/12}.
\end{equation}
The relation between $\mu$, $I$ and $\dot\gamma^*$ can be obtained combining \Eqs\eqref{nuIstar} and \eqref{mustar}, and involves also the volume fraction. \Fig\ref{muI_figs} confirms the good agreement of the model with the data. The discrepancies in the stress ratio at large inertial number, $I > 0.3$, correspond to those at $0.2 \leq \nu \leq 0.6$ in \Fig\ref{mu_fig}(a), while the relation between $I$ and density is captured well throughout.

\section{Comparison with other models}

In this Section, we compare our model with other theories developed for deformable particles, from the literature. Here we do neither consider rheologies which apply only to rigid particles, such as the $\mu-I$ model or similar,\cite{gdr2004,dac2005,jop2006,pey2008} nor standard kinetic theory based models.\cite{gar1999,ves2013,ves2014}
In particular, we focus on the constitutive models proposed by \citet{chi2012}, \citet{ber2015b} and \citet{sin2015}
Moreover, we consider also the model proposed by \citet{par2013} derived for emulsion-like systems.
In the following, we present the four models and compare them with ours and the numerical data.

\begin{enumerate}

\item \citet{chi2012} proposed a new definition of the scaled pressure and the macroscopic friction:
\begin{align}
p^{*} &= \begin{cases}
\left( p^{*-1}_{\inert} + p^{*-1}_{\inte} \right)^{-1}, & \mbox{if } \nu \leq \nu_\jam,\\
p^*_{\qs} + p^*_{\inte}, & \mbox{if } \nu > \nu_\jam,
\end{cases}\label{chialvo_pstar}\\
\mu &= \mu_{\hard} - \mu_{\soft},\label{chialvo_mu}
\end{align}
where:
\[
\begin{array}{ll}
p^{*}_{\inert} =a_1\dfrac{\dot\gamma^{*2}}{\left(\nu_\jam - \nu\right)^2}, & 
p^{*}_{\qs} =a_2\left(\nu-\nu_\jam\right)^{2/3}\\
p^{*}_{\inte} =a_3 \dot\gamma^{*1/2}, & 
\mu_{\soft} = \dfrac{a_4}{\dot\gamma_0^*/\dot\gamma^*+1}.
\end{array}
\]
$p^{*}_{\inert}$, $p^{*}_{\qs}$ and $p^{*}_{\inte}$ define the scaled pressure in the three regimes: fluid (named inertial in Ref.~\cite{chi2012}), solid (quasi-static) and intermediate, respectively; $\mu_{\soft}$ is the correction to the standard $\mu-I$ rheology to account for softness effects; $a_1 = 0.021$, $a_2 = 0.095$, $a_3 = 0.099$, $a_4 = 0.2$ and $\dot\gamma_0^* = 0.1$ are dimensionless model parameters. 
Furthermore, for systems of frictionless, monodispersed particles, the authors in Ref.~\cite{chi2012} estimated the jamming volume fraction to be 0.636.
The standard $\mu-I$ rheology is here denoted as $\mu_{\hard}$ and given by 
\begin{equation}\label{muI}
\mu_{\hard} = \mu_0 + \dfrac{\left(\mu_\infty-\mu_0\right)}{I_0/I+1}
\end{equation}
with $I = \dot\gamma^*/p^{*1/2}$ and $\mu_0 = 0.12$, $\mu_\infty = 0.55$, $I_0 = 0.2$ for frictionless particles. The same expression for $p^*$ has been adopted by Ness and Sun in Ref.~\cite{nes2015}, where the authors also add a stress contribution for the intermediate viscous fluid in order to extend the rheology to non-Brownian suspensions. \\

\begin{figure}[!h]
\centering
\includegraphics[width=1.\textwidth]{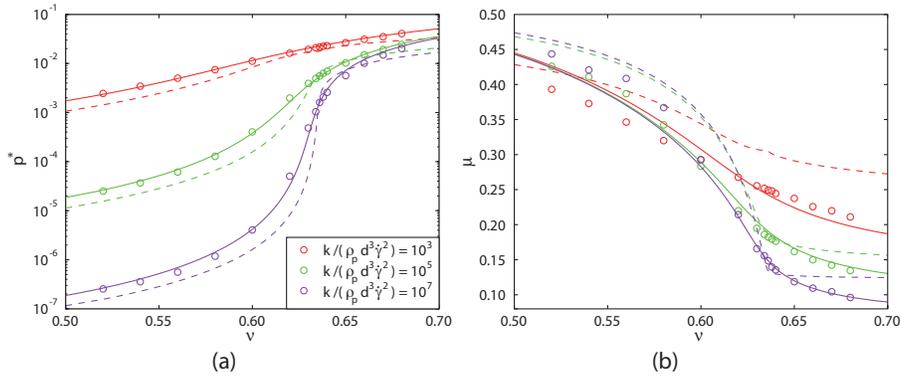}
\caption{Scaled pressure $p^*$ (a) and stress ratio $\mu$ (b) plotted against the volume fraction, for the proposed constitutive model (solid lines, \Eqs\eqref{pstar} and \eqref{mustar}), and the constitutive model of \citet{chi2012} (dashed lines, \Eqs\eqref{chialvo_pstar} and \eqref{chialvo_mu}). Symbols represent our numerical data.}
\label{chialvo_comparisons}
\end{figure}

\noindent \Fig\ref{chialvo_comparisons} depicts, for three cases $k/\left(\rho_p d^3 \dot\gamma^2\right) = 10^3$, $10^5$ and $10^7$, the comparison of the present model (solid lines) with the theory of \citet{chi2012} (dashed lines), which underpredicts the pressure and overpredicts the stress ratio in both fluid and solid regimes. Besides the shift, the trend of the stress ratio at small volume fractions, $\nu < 0.6$, is qualitatively better captured by the model of Chialvo et al., which predicts the slight rate dependency of $\mu$ even in the fluid regime (\Fig\ref{chialvo_comparisons}b).\\
Concerning $p^*$, the bad prediction of \Eq\eqref{chialvo_pstar} is mainly due to the powers of $\left|\nu-\nu_\jam\right|$ which govern the fluid and solid response, which are $-2$ and $2/3$, respectively. 
If we consider only the fluid and solid regimes in the theory of Chiavo et al., $p^*_{\inert}$ and $p^*_{\qs}$, then we can merge the two equations using our approach:
\begin{equation}\label{pstar2}
\nu - \nu_\jam = \left(\dfrac{p^*}{a_2}\right)^{3/2} - \left(\dfrac{a_1\dot\gamma^{*2}}{p^*}\right)^{1/2}.
\end{equation}
The curves obtained using \Eq\eqref{pstar2} completely overlap with those of Chialvo \emph{et al}. in \Fig\ref{chialvo_comparisons}(a) (note that the comparison is not shown in the figure for the sake of clarity) meaning that the two approaches result in very similiar predictions for $p^*$.
The form of \Eq\eqref{pstar2} is simpler than \Eq\eqref{chialvo_pstar}, because it (i) does not need the definition of the intermediate regime pressure and (ii) does not require an ``if'' condition to distinguish below and above the jamming volume fraction, given that any divergence is avoided.  
Anyway, the critical exponents adopted in our model (-12/5 for the fluid and 6/5 for the solid regime), together with the extra dependence of $p^*$ on $\nu$ and the choice $\nu_\jam = 0.634$, allow a better quantitative agreement with most of our numerical data.
Note however, that we did not fit the parameters of Chialvo et al. to our data, so that the slight shift must be disregarded, while the qualitative behavior can be appreciated.\\

\item \citet{ber2015b} have extended the standard kinetic theory to account for the deformability of the particles. Their model can be summarized as follows:
\begin{align}
p^{*} &= \begin{cases}
\left( p^{*-1}_{\rigi} + p^{*-1}_{\defo} \right)^{-1}, & \mbox{if } \nu \leq \nu_\jam,\\
p^*_{\el} + p^*_{\defo}, & \mbox{if } \nu > \nu_\jam,
\end{cases}\label{berzi_pstar}\\
s^{*} &= \begin{cases}
\left( s^{*-1}_{\rigi} + s^{*-1}_{\defo} \right)^{-1}, & \mbox{if } \nu \leq \nu_\jam,\\
s^*_{\el} + s^*_{\defo}, & \mbox{if } \nu > \nu_\jam.
\end{cases}\label{berzi_sstar}
\end{align}
Similarly to \Eq\eqref{chialvo_pstar}, scaled pressure and shear stress are given by using three contributions: rigid (subscript rig), deformable (subscript def) and elastic (subscript el):
\[
\begin{array}{lll}
p^*_{\rigi} = f_1 T^*, & p^*_{\el} = b_1 \left(\nu-\nu_\jam\right), & p^*_{\defo} = b_2 \nu T^{*1/2}, \\
s^*_{\rigi} = f_2 T^{*1/2}\dot\gamma^*, & s^*_{\el} = b_3 \ p^*_{el}, & s^*_{\defo} = b_4 \nu \dot\gamma^*.
\end{array}
\]
Here, $b_1 = 0.6$, $b_2 = 1.42$, $b_3 = 0.11$, $b_4 = 0.36$ for the case of frictionless particles having $e_n = 0.7$, whereas $f_1$ and $f_2$ are functions of the volume fraction and the coefficient of restitution, derived from kinetic theory, and summarized in \Tab\ref{tab2}. For frictionless particles, the authors have used $\nu_\jam = 0.636$.
The scaled granular temperature $T^*$ is computed as solution to the balance of fluctuation energy and, in simple shear conditions, results in:
\begin{equation}\label{Tstar_berzi}
T^{*} = \begin{cases}
b_5\left[1+b_6\dfrac{\max\left(\nu-\nu_f,0\right)}{\nu_{rcp}-\nu}\right]\dot\gamma^{*2}, & \mbox{if } \nu \leq \nu_\jam,\\
b_7\dot\gamma^{*2}, & \mbox{if } \nu > \nu_\jam
\end{cases}
\end{equation}
with $\nu_f = 0.49$, $\nu_{rcp}=0.64$, $b_5 = 0.25$, $b_6 = 0.52$ and $b_7 = 5.06$. 

\begin{table}[!h]
\begin{center}
\caption{List of functions in the constitutive relations of \citet{ber2015b}}
\label{tab2}
\begin{tabular}{l}
\hline
$f_1 = 2(1+e_n)g_0\nu^2$\\
$f_2 = 8Jg_0\nu^2/\left(5\pi^{1/2}\right)$\\
$J = (1+e_n)/2 + \pi\left(1+e_n\right)^2(3e_n-1)/\left[96-24\left(1-e_n\right)^2-20\left(1-e_n^2\right)\right]$\\
$g_0 = f(2-\nu)/2/\left(1-\nu\right)^3 + 2(1-f)/(\nu_\jam-\nu)$\\
$f = (\nu_\jam+\nu-0.8)(\nu_\jam-\nu)/\left(\nu_\jam-0.4\right)^2$ \\
\hline
\end{tabular}
\end{center}
\end{table}

\begin{figure}[!h]
\centering
\includegraphics[width=1.\textwidth]{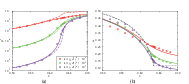}
\caption{Scaled pressure $p^*$ (a) and stress ratio $\mu$ (b) plotted against the volume fraction, for the proposed constitutive model (solid lines, \Eqs\eqref{pstar} and \eqref{mustar}), and the constitutive model of \citet{ber2015b} (dash-dotted lines, \Eq\eqref{berzi_pstar} and ratio of \Eqs\eqref{berzi_sstar} to \eqref{berzi_pstar}). Symbols represent our numerical data.}
\label{berzi_comparisons}
\end{figure}

\noindent In \Fig\ref{berzi_comparisons} we show the comparison of our model (solid lines) with that of Berzi and Jenkins (dash-dotted lines). For the latter, the stress ratio is computed just dividing \Eq\eqref{berzi_sstar} by \Eq\eqref{berzi_pstar}. The major limitation of the theory in Ref.~\cite{ber2015b} is that the stress ratio is not affected by the particle stiffness (\Fig\ref{berzi_comparisons}b). This is due to the assumption that the scaled granular temperature scales with $\dot\gamma^{*2}$ even at $\nu > \nu_\jam$, \Eq\eqref{Tstar_berzi}; this assumption, besides being in disagreement with what is shown by the numerical data (\Fig\ref{T_fig}a), makes the stress ratio constant in the solid regime:
\[
\mu = \dfrac{s^*_{\el}+s^*_{\defo}}{p^*_{\el}+p^*_{\defo}} = b_3,
\]
given that the model parameters, in simple shearing, are such that $b_4 = b_2 b_3 b_7^{1/2}$.
On the other hand, it must be noticed that the theory of Berzi and Jenkins is generally developed for any kind of flow configurations, being described by conservation laws of mass, momentum and fluctuation energy and a full, more complete set of constitutive relations, including also the coefficient of restitution. Moreover, at very low densities, $\nu < 0.2$, the model in Ref.~\cite{ber2015b} corresponds to the standard kinetic theory in Ref.~\cite{gar1999} which has been proven to quantitatively predict all the variables, and, in particular the shear stress, differently from the model proposed here.

\item \citet{sin2015}, as rephrased in Ref.~\cite{lud2016c,lud2016b}, have generalized the standard $\mu-I$ relation accounting for the influence of the particle stiffness through the scaled pressure $p^*$:
\begin{align}
\mu &= \mu_{\hard}\left[1-\left(\dfrac{p^*}{p^*_0}\right)^{1/2}\right], \label{luding_mu}\\
\nu &= \nu_c\left(1-\dfrac{I\sqrt{\nu}}{I_{0\nu}}\right)\left(1+\dfrac{p^*}{p^*_{0\nu}}\right), \label{luding_nu}
\end{align}
with $\mu_{\hard}$ given by \Eq\eqref{muI}, and constant parameters $p^*_0 = 0.9$, $I_{0\nu} = 0.85$ and $p^*_{0\nu}=0.33$ and $\nu_c = 0.642$. 
Note that the data in Ref.~\cite{sin2015} have been obtained using the ring-shear geometry (inhomogeneous), by local coarse-graining,  and slightly frictional and polydisperse particles, which implies a small shift in $\nu_c$ and some other parameters. In particular, in order to compare our results with the model of Singh et al., here we adopt $I_{0\nu} = 3.28$ and $\nu_c = 0.634$ suitable for monodisperse systems.
Furthermore, the additive corrections in Ref.~\cite{sin2015} have been rewritten (identical in first order approximation) as multiplicative correction terms in Ref.~\cite{lud2016b} that imply an higher order non-linear correction neglected in Ref.~\cite{sin2015}.
Using \Eq\eqref{luding_nu}, $p^*$ can be computed implicitly as function of $\nu$ and $\dot\gamma^*$ (or, equivalently, $k$). The present theory is compared with the model of Singh et al. in \Fig\ref{luding_comparisons}. \\
%
\begin{figure}[!h]
\centering
\includegraphics[width=1.\textwidth]{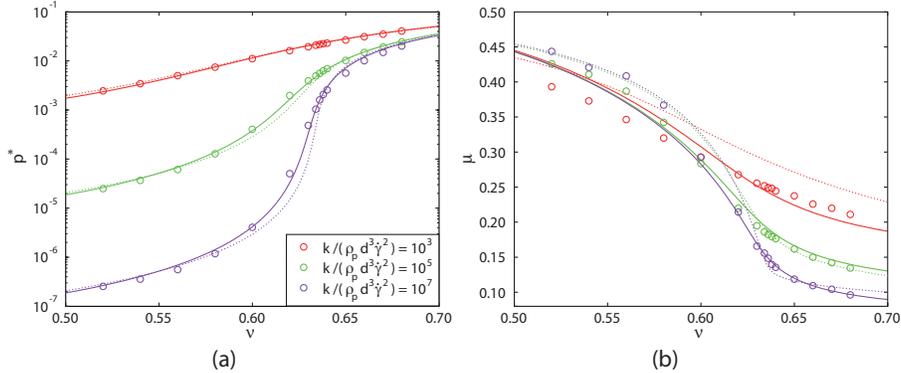}
\caption{Scaled pressure $p^*$ (a) and stress ratio $\mu$ (b) plotted against the volume fraction, for the proposed constitutive model (solid lines, \Eqs\eqref{pstar} and \eqref{mustar}), and the constitutive model of \citet{sin2015} (dotted lines, \Eqs\eqref{luding_nu} and \eqref{luding_mu}). Symbols represent our numerical data.}
\label{luding_comparisons}
\end{figure}
%
\noindent As it has been already stated, we have calibrated the parameters in the model of Singh et al. by using our numerical results, as a consequence the model results in a good quantitative agreement with the data, except for very soft particles ($k/\left(\rho_p d^3 \dot\gamma^2\right)=10^3$).
As for the model of \citet{chi2012}, the stress ratio predicted by the model of Singh et al. is able to qualitatively capture the rate dependency of $\mu$ at $\nu < \nu_\jam$. 

\item Finally, \citet{par2013} have presented a microscopic two-state theory for yield-stress fluids, elaborated in more detail by \citet{din2015}, to describe the transition between jammed and unjammed states. Yield-stress fluids are complex fluids composed of dispersions of one material (particles, drops or bubbles) in a liquid (or continuous phase), whose mechanical behavior is characterized by the emergence of a yield stress for volume fractions higher than some critical value (the jamming volume fraction) and Newtonian flow for lower volume fractions, with shear thinning in either case for high shear rates. The measurements of the shear stress obtained in the experiments of \citet{par2013} and \citet{din2015} with different kinds of yield-stress fluids, collapse when rescaling all data as: $s^*/\left|\nu - \nu_\jam \right|^{\alpha}$ - $\dot\gamma^*/\left|\nu - \nu_\jam \right|^{\beta}$, with $\alpha$ and $\beta$ scaling parameters, similar to our \Fig\ref{scaling_fig}(b). According to this collapse, the scaled shear stress has been found to obey: (i) the Herschel-Bulkley \cite{her1926} equation when $\nu > \nu_\jam$, with yield stress expressed as a power law in the distance to jamming, and (ii) the Cross equation \cite{cro1965} when $\nu < \nu_\jam$, where the Newtonian viscosity satisfies a power law in $\left|\nu - \nu_\jam \right|$. This result can be summarized as
\begin{equation}\label{eq_paredes}
s^{*} = \begin{cases}
\eta_0 \dfrac{\dot\gamma^{*m^{\fluid}}}{\left|\nu - \nu_\jam \right|^{m^{\fluid}\beta-\alpha}}\left[ 1+\eta_0\dfrac{\dot\gamma^{*(m^{\fluid}-\alpha/\beta)}}{s_1 \left|\nu - \nu_\jam \right|^{m^{\fluid}\beta-\alpha}} \right]^{-1}, & \mbox{if } \nu < \nu_\jam,\\
s_0\left|\nu - \nu_\jam \right|^\alpha + s_1\dot\gamma^{*\alpha/\beta}, & \mbox{if } \nu \geq \nu_\jam,
\end{cases}
\end{equation}
where $s_0$, $s_1$ and $\eta_0$ are adjustable dimensionless parameters and the coefficients $\alpha$, $\beta$ and $m^{\fluid}$ are given in \Fig\ref{paredes_comparisons}. The yield stress and the Newtonian viscosity are given by $s_0\left|\nu - \nu_\jam \right|^\alpha$ and $\eta_0 \dot\gamma^{*m^{\fluid}}/\left|\nu - \nu_\jam \right|^{m^{\fluid}\beta-\alpha}$, respectively.\\
As previously stated, the critical exponents, as well as the jamming volume fraction, are material-dependent.
The original model of Paredes et al. has been derived for soft matter systems, and although it is compatible in spirit with the granular rheology, is completely different from our granular fluid in all quantitative numbers and exponents, especially below jamming.
As a consequence, it requires careful calibration of the parameters, differently from the models previously analyzed, specifically derived for granular materials.
In particular, the exponent $m^{\fluid}$, appearing in the unjammed phase, is measured equal to 1 for any kind of yield-stress fluid considered. 
$m^{\fluid} = 1$ implies that, in the fluid regime, far from jamming, the shear stress varies linearly with the shear rate. 
This linear dependence, typical of viscous liquids, does not apply to dry granular systems, where $s^* \propto \dot\gamma^{*2}$ (Bagnold scaling), implying $m^\fluid = 2$. 
In order to adapt the model of Paredes et al. to dry granular systems, we use $m^\fluid = 2$ in \Eq\eqref{eq_paredes} and calibrate the critical exponents and the dimensionless parameters of the model by collapsing our numerical data on the plane $s^*/\left|\nu - \nu_\jam \right|^{\alpha}$ - $\dot\gamma^*/\left|\nu - \nu_\jam \right|^{\beta}$, \Fig\ref{paredes_comparisons}. 
We obtain $\nu_\jam = 0.634$, $\alpha = 6/5$ and $\beta = 16/5$
(whereas \citet{din2015} have estimated, for different kinds of yield-stress fluids, $\nu_\jam = 0.64\div 0.68$, $\alpha = 2.04\div 2.21$ and $\beta = 3.75\div 3.84$).
These values coincide with those reported in \Tab\ref{tab1} for $s^*$, in fact \Fig\ref{paredes_comparisons} corresponds to \Fig\ref{scaling_fig}(b) unless for $\nu^{1/2}$ (actually this last provides a better collapse of the data in the unjammed regime). 
In \Fig\ref{paredes_comparisons}, \Eq\eqref{eq_paredes} is also plotted: the dashed line represents the Herschel-Bulkley (branch at $\nu \geq \nu_\jam$) and dash-dotted line represents the Cross equation ($\nu < \nu_\jam$) with $m^\fluid = 2$. 
By fitting the data, we infer the dimensionless parameters appearing in \Eq\eqref{eq_paredes}: $s_0 = 0.08$, $s_1 =0.07$ and $\eta_0 = 0.015$.
Note that both branches of \Eq\eqref{eq_paredes} are defined in proximity of jamming $\nu \sim \nu_\jam$ and fit well the data in this region. Moreover, both the Herschel-Bulkley equation and the Cross equation lead to the same expression $s^* = s_1 \dot\gamma^{*\alpha/\beta} = s_1 \dot\gamma^{*3/4}$ at the jamming transition $\nu \longrightarrow \nu_\jam$.\\
%
\begin{figure}[!h]
\centering
\includegraphics[width=1.\textwidth]{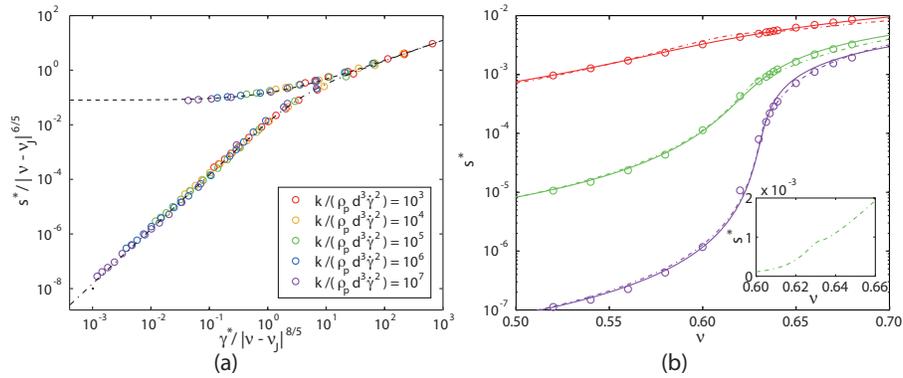}
\caption{(a) Collapse of scaled shear stress plotted against scaled shear rate as suggested in Ref.~\cite{par2013}, for different values of the
(dimensionless) particle stiffness. Dash-dotted line ($\nu < \nu_\jam$) and dashed line ($\nu \geq \nu_\jam$) represent the two branches of \Eq\eqref{eq_paredes} where $\alpha = 6/5$, $\beta = 16/5$, $m^\fluid = 2$, $s_0 = 0.08$, $s_1 =0.07$ and $\eta_0 = 0.015$. (b) Scaled shear stress plotted against the volume fraction, for the proposed constitutive model (solid lines, \Eq\eqref{sstar}), and the model of \citet{par2013} (dash-dash-dotted lines, \Eq\eqref{eq_paredes}). Symbols represent our numerical data. The inset shows the scaled shear stress predicted by \Eq\eqref{eq_paredes} for the case $k/\left(\rho_p d^3 \dot\gamma^2\right) = 10^5$, plotted in linear scale, at volume fractions close to jamming.}
\label{paredes_comparisons}
\end{figure}
%
\noindent In \Fig\ref{paredes_comparisons}(b) the scaled shear stress obtained with the present theory, \Eq\eqref{sstar}, is compared with the model of Paredes et al., \Eq\eqref{eq_paredes}. Since the parameters in the model of Paredes et al. have been calibrated on the basis of our numerical data, the agreement is very good. 
Nevertheless, due to the switching conditions in the proximity of jamming, \Eq\eqref{eq_paredes} shows a wiggle at $\nu = \nu_\jam$, as depicted in the inset of \Fig\ref{paredes_comparisons}(b) (where the linear scale is adopted instead of the logaritmic one), being a function continuous but not differentiable  with respect to $\nu$ at the jamming volume fraction.

\end{enumerate}

\section{Conclusions}

In this paper, we have performed DEM simulations of simple shear flows of granular materials composed of frictionless, deformable spheres, in order to investigate the role of particles stiffness. A wide range of volume fractions has been analyzed to cover the entire set of granular flow regimes, that is from fluid to solid conditions, including the transition between the two states. 
The main goal is to propose a phenomenological, constitutive model, based on continuously and differentiably implemented equations, which accounts for the particles' softness over orders of magnitude, and predicts stresses and granular temperature at all densities. 
The constitutive relations are given by functions which smoothly merge the scaling relations in the fluid and solid limits with shear rate and distance to jamming, as control parameters. 
Before presenting the merging functions, we have inferred the asymptotical scaling relations in the two regimes by separately collapsing our DEM results for all stiffnesses, and obtained critical exponents for the distance from jamming slightly different to those reported in other works. \\
In particular, we have found the shear stress to be weakly shear rate dependent in the solid regime, varying with a rather small power 1/6. This rate dependency, for the wide range of particle stiffnesses investigated, has been confirmed by plotting the macroscopic friction $\mu$ (ratio between the shear stress and the pressure, which is rate-independent) versus the volume fraction, for which the data do not collapse. As a consequence, given that in the solid regime the scaled shear rate affects the stress ratio, the shear stress cannot be rate independent as the pressure is. 
Nevertheless, due to the lack of numerical data at very large particle stiffness (or, equivalently, scaled shear rates), in the solid regime, we cannot conclude on the functional behavior of the shear stress rate dependence; as a consequence, the constitutive relation must be revised in future in order to deal with a wider range of stiffnesses.\\
Merging is done by stitching the two scaling regimes implicitly.
Independently of the scaling exponents and the rate dependency of the shear rate, the form of the proposed merged constitutive relations has the advantage that it does not require the definition of additional merging functions in the transitional regime near to the jamming volume fraction. Moreover, it does not include any divergent quantity at the jamming volume fraction, and, as a consequence, can be continuously implemented without any ``if'' or switching condition in the proximity of jamming. 
The model can be easily adapted to merge any kind of relations describing the fluid and the solid regimes, provided that both are expressed as power-law relations of the distance to jamming.
The model has been shown to quantitatively predict all flow variables in a wide range of densities, spanning from dilute (fluid regime) to dense (solid regime) conditions, including the transitional regime. The comparison with other models from literature has illustrated a better prediction of the quantities, especially in the solid regime. 
The merged model does not require additional fitting parameters, but only constant material parameters, inferred from the numerical simulations in the fluid and solid limits, respectively. 
We expect these material parameters to be affected by other material properties, such as dissipation (via the coefficient of restitution) and inter-particle friction in case of frictional particles. Calibration of the constitutive model for other coefficients of restitution and non-zero friction will be the subject of future work.
Future research can also focus on appropriate calibration of the model parameters to deal with wet particulate systems\cite{nes2015} and soft matter,\cite{par2013} such as emulsions, colloids, foams, gels and suspensions of (soft) particles, which exhibit a jamming transition and scaling behaviors similar to dry granular systems.
Finally, in the limit of very dilute systems, i.e., volume fractions smaller than 0.4, the present model is not very accurate. In such conditions, the kinetic theory of granular gases has been demonstrated to successfully predict all the quantities. An improvement of the current model would be to merge the relations in the lower-density fluid regime with the kinetic theory.

\section*{Acknowledgements}

This work was financially supported by the NWO/STW VICI grant 10828. Dalila Vescovi is supported by a fellowship from Fondazione Fratelli Confalonieri.

\begin{appendices}

\section*{Appendix A}

In this Appendix, we provide the numerical data obtained at the steady state in our DEM simulations.
In particular, the volume fraction, $\nu$, dimensionless pressure, $p/\left(\rho_p d^2 \dot\gamma^{2}\right)$, shear stress, $s/\left(\rho_p d^2 \dot\gamma^{2}\right)$, granular temperature, $T/\left(d^2 \dot\gamma^{2}\right)$ and coordination number $C$ are reported for five values of the dimensionless particle stiffness $k/\left(\rho_p d^3 \dot\gamma^2\right)$: $10^3$, $10^4$, $10^5$, $10^6$ and $10^7$ (\Tab\ref{tab1}-\ref{tab5}, respectively). In the simulations, we have used normal coefficient of restitution $e_n = 0.7$, tangential coefficient of restitution $e_t = -1$, interparticle friction coefficient $\mu = 0$ and normal spring stiffness $k_n = k$.

\begin{table}[h]
\begin{center}
\caption{\ Summary of measured steady state quantities for $k/\left(\rho_p d^3 \dot\gamma^2\right) = 10^3$}
\label{tab1}
\begin{tabular}{ccccc}
\hline
$\nu$ & $p/\left(\rho_p d^2 \dot\gamma^{2}\right)$ & $s/\left(\rho_p d^2 \dot\gamma^{2}\right)$ & $T/\left(d^2 \dot\gamma^{2}\right)$ & $C$\\
\hline
0.200 & 0.234942 & 0.103128 & 0.555758 & 0.166381\\
0.250 & 0.265588 & 0.117213 & 0.394930 & 0.209494\\
0.300 & 0.331448 & 0.148527 & 0.314988 & 0.273373\\
0.350 & 0.445152 & 0.200126 & 0.275619 & 0.369064\\
0.400 & 0.642769 & 0.285643 & 0.258288 & 0.519074\\
0.450 & 1.023252 & 0.443855 & 0.265733 & 0.783821\\
0.500 & 1.839065 & 0.749744 & 0.293879 & 1.261998\\
0.520 & 2.457556 & 0.966064 & 0.319781 & 1.568773\\
0.540 & 3.437453 & 1.281308 & 0.359457 & 1.981529\\
0.560 & 4.981236 & 1.726183 & 0.415063 & 2.506194\\
0.580 & 7.445717 & 2.381977 & 0.490317 & 3.140327\\
0.600 & 11.140037 & 3.260685 & 0.580537 & 3.829470\\
0.620 & 16.326774 & 4.371615 & 0.681840 & 4.518889\\
0.630 & 19.529057 & 4.989191 & 0.732009 & 4.853679\\
0.634 & 20.893332 & 5.264431 & 0.752643 & 4.979053\\
0.636 & 21.603690 & 5.388448 & 0.762954 & 5.045281\\
0.638 & 22.325605 & 5.538356 & 0.772385 & 5.101963\\
0.640 & 23.078818 & 5.642747 & 0.787518 & 5.163304\\
0.650 & 26.592465 & 6.315878 & 0.830628 & 5.478336\\
0.660 & 31.206236 & 7.049504 & 0.878267 & 5.750006\\
0.670 & 35.389524 & 7.785759 & 0.918761 & 6.044560\\
0.680 & 40.693236 & 8.589171 & 0.959013 & 6.274154\\
\hline
\end{tabular}
\end{center}
\end{table}

\begin{table}[h]
\begin{center}
\caption{\ Summary of measured steady state quantities for $k/\left(\rho_p d^3 \dot\gamma^2\right) = 10^4$}
\label{tab2}
\begin{tabular}{ccccc}
\hline
$\nu$ & $p/\left(\rho_p d^2 \dot\gamma^{2}\right)$ & $s/\left(\rho_p d^2 \dot\gamma^{2}\right)$ & $T/\left(d^2 \dot\gamma^{2}\right)$ & $C$\\
\hline
0.200 & 0.239205 & 0.106725 & 0.550503 & 0.054813\\
0.250 & 0.276281 & 0.122196 & 0.393941 & 0.070244\\
0.300 & 0.348642 & 0.158886 & 0.315481 & 0.092023\\
0.350 & 0.465862 & 0.215828 & 0.275625 & 0.124473\\
0.400 & 0.676403 & 0.309074 & 0.254802 & 0.180054\\
0.450 & 1.060903 & 0.471786 & 0.254966 & 0.279676\\
0.500 & 1.911529 & 0.826375 & 0.270529 & 0.486684\\
0.520 & 2.615479 & 1.088301 & 0.291918 & 0.647978\\
0.540 & 3.831324 & 1.520142 & 0.328693 & 0.903100\\
0.560 & 6.437990 & 2.343001 & 0.399776 & 1.352436\\
0.580 & 12.523880 & 4.090963 & 0.536589 & 2.096784\\
0.600 & 28.197296 & 7.864387 & 0.802913 & 3.187979\\
0.620 & 63.395662 & 15.028613 & 1.219510 & 4.366442\\
0.630 & 90.035044 & 19.745010 & 1.456724 & 4.891566\\
0.634 & 102.666463 & 22.121499 & 1.553021 & 5.085384\\
0.636 & 109.179503 & 23.286849 & 1.608156 & 5.178923\\
0.638 & 116.084014 & 24.140719 & 1.651068 & 5.272760\\
0.640 & 122.769207 & 25.198054 & 1.705943 & 5.355511\\
0.650 & 157.343054 & 30.582846 & 1.924566 & 5.782143\\
0.660 & 203.511144 & 37.024106 & 2.148040 & 6.133889\\
0.670 & 245.825537 & 43.124529 & 2.349487 & 6.485831\\
0.680 & 300.964068 & 49.547237 & 2.521318 & 6.756794\\
\hline
\end{tabular}
\end{center}
\end{table}

\begin{table}[h]
\begin{center}
\caption{\ Summary of measured steady state quantities for $k/\left(\rho_p d^3 \dot\gamma^2\right) = 10^5$}
\label{tab3}
\begin{tabular}{ccccc}
\hline
$\nu$ & $p/\left(\rho_p d^2 \dot\gamma^{2}\right)$ & $s/\left(\rho_p d^2 \dot\gamma^{2}\right)$ & $T/\left(d^2 \dot\gamma^{2}\right)$ & $C$\\
\hline
0.200 & 0.241428 & 0.108570 & 0.543630 & 0.017753\\
0.250 & 0.278552 & 0.126415 & 0.392905 & 0.022594\\
0.300 & 0.351964 & 0.161171 & 0.314589 & 0.029670\\
0.350 & 0.477694 & 0.219933 & 0.273076 & 0.040755\\
0.400 & 0.685554 & 0.318180 & 0.252738 & 0.058722\\
0.450 & 1.081814 & 0.491962 & 0.248999 & 0.093042\\
0.500 & 1.864913 & 0.824706 & 0.255639 & 0.162584\\
0.520 & 2.513597 & 1.071019 & 0.268573 & 0.221985\\
0.540 & 3.656958 & 1.502096 & 0.293246 & 0.326832\\
0.560 & 6.085880 & 2.354844 & 0.344352 & 0.541056\\
0.580 & 12.784651 & 4.375863 & 0.462589 & 1.021208\\
0.600 & 40.463774 & 11.463297 & 0.795484 & 2.139451\\
0.620 & 196.788428 & 43.262171 & 1.867153 & 4.023546\\
0.630 & 396.324077 & 77.243801 & 2.788652 & 4.856156\\
0.634 & 495.184615 & 92.030928 & 3.163740 & 5.065659\\
0.636 & 555.923377 & 101.259968 & 3.361877 & 5.222647\\
0.638 & 629.920288 & 113.051389 & 3.614110 & 5.411202\\
0.640 & 693.979225 & 122.525143 & 3.820721 & 5.523332\\
0.650 & 1026.926498 & 166.670169 & 4.718021 & 6.047363\\
0.660 & 1487.286922 & 222.807657 & 5.682130 & 6.447426\\
0.670 & 1920.393069 & 273.270325 & 6.338657 & 6.842727\\
0.680 & 2436.413902 & 327.387775 & 7.004498 & 7.155652\\
\hline
\end{tabular}
\end{center}
\end{table}

\begin{table}[h]
\begin{center}
\caption{\ Summary of measured steady state quantities for $k/\left(\rho_p d^3 \dot\gamma^2\right) = 10^6$}
\label{tab4}
\begin{tabular}{ccccc}
\hline
$\nu$ & $p/\left(\rho_p d^2 \dot\gamma^{2}\right)$ & $s/\left(\rho_p d^2 \dot\gamma^{2}\right)$ & $T/\left(d^2 \dot\gamma^{2}\right)$ & $C$\\
\hline
0.200 & 0.240997 & 0.110712 & 0.542786 & 0.005486\\
0.250 & 0.285286 & 0.127245 & 0.393289 & 0.007462\\
0.300 & 0.347973 & 0.158695 & 0.314593 & 0.009479\\
0.350 & 0.487812 & 0.220737 & 0.272417 & 0.013305\\
0.400 & 0.701025 & 0.328384 & 0.252236 & 0.018991\\
0.450 & 1.054117 & 0.504345 & 0.246853 & 0.028296\\
0.500 & 1.835065 & 0.836030 & 0.249521 & 0.051556\\
0.520 & 2.459094 & 1.061434 & 0.259626 & 0.071754\\
0.540 & 3.539649 & 1.464188 & 0.278913 & 0.108506\\
0.560 & 5.766162 & 2.287424 & 0.317835 & 0.187553\\
0.580 & 12.055349 & 4.214920 & 0.413838 & 0.409660\\
0.600 & 41.481005 & 12.137577 & 0.701306 & 1.148615\\
0.605 & 65.253233 & 17.711560 & 0.875353 & 1.534955\\
0.610 & 106.876056 & 27.037876 & 1.120896 & 2.001262\\
0.615 & 197.302649 & 46.046854 & 1.523431 & 2.631728\\
0.620 & 391.976552 & 84.007070 & 2.208526 & 3.358247\\
0.625 & 803.442184 & 154.964778 & 3.311855 & 4.097119\\
0.630 & 1585.129284 & 279.067958 & 4.980820 & 4.769387\\
0.634 & 2403.201354 & 399.027991 & 6.289582 & 5.164167\\
0.635 & 2684.443882 & 437.334838 & 6.885728 & 5.258285\\
0.636 & 2926.634561 & 475.040850 & 7.149895 & 5.348844\\
0.638 & 3560.018222 & 547.873183 & 8.044064 & 5.530090\\
0.640 & 4123.181555 & 634.939118 & 8.850524 & 5.660419\\
0.650 & 7370.295715 & 1037.449061 & 12.455813 & 6.261044\\
0.660 & 11680.402608 & 1483.608749 & 15.902378 & 6.716673\\
0.670 & 16435.452806 & 1964.286219 & 18.502284 & 7.076342\\
0.680 & 21702.678475 & 2401.395598 & 21.750145 & 7.379358\\
\hline
\end{tabular}
\end{center}
\end{table}

\begin{table}[h]
\begin{center}
\caption{\ Summary of measured steady state quantities for $k/\left(\rho_p d^3 \dot\gamma^2\right) = 10^7$}
\label{tab5}
\begin{tabular}{ccccc}
\hline
$\nu$ & $p/\left(\rho_p d^2 \dot\gamma^{2}\right)$ & $s/\left(\rho_p d^2 \dot\gamma^{2}\right)$ & $T/\left(d^2 \dot\gamma^{2}\right)$ & $C$\\
\hline
0.200 & 0.242585 & 0.102214 & 0.545485 & 0.001765\\
0.250 & 0.283350 & 0.128201 & 0.392078 & 0.002291\\
0.300 & 0.340522 & 0.158116 & 0.314910 & 0.002822\\
0.350 & 0.465374 & 0.205602 & 0.273811 & 0.003982\\
0.400 & 0.699502 & 0.331268 & 0.250779 & 0.005899\\
0.450 & 1.120364 & 0.500994 & 0.245302 & 0.010071\\
0.500 & 1.840210 & 0.839447 & 0.248525 & 0.017071\\
0.520 & 2.532813 & 1.123175 & 0.256162 & 0.023502\\
0.540 & 3.544851 & 1.490749 & 0.273729 & 0.035264\\
0.560 & 5.584651 & 2.280780 & 0.309487 & 0.061977\\
0.580 & 11.767423 & 4.319858 & 0.392010 & 0.149072\\
0.600 & 40.479041 & 11.845031 & 0.658790 & 0.530188\\
0.620 & 504.648534 & 108.158916 & 2.153449 & 2.434020\\
0.630 & 4824.238310 & 799.621737 & 7.571104 & 4.504102\\
0.634 & 10305.288288 & 1581.602954 & 11.858220 & 5.097600\\
0.636 & 15082.792004 & 2183.362941 & 15.798637 & 5.378536\\
0.638 & 20808.342189 & 2901.934944 & 19.518591 & 5.612210\\
0.640 & 25657.006376 & 3473.088378 & 22.408651 & 5.763377\\
0.650 & 58191.181725 & 7142.838075 & 36.532318 & 6.433599\\
0.660 & 101401.733582 & 11086.571932 & 50.484634 & 6.880554\\
0.670 & 148887.949649 & 15539.824114 & 57.148331 & 7.238029\\
0.680 & 202098.496818 & 19406.096112 & 72.691066 & 7.530354\\
\hline
\end{tabular}
\end{center}
\end{table}

\end{appendices}

\clearpage

\bibliographystyle{plainnat}
\normalem

\end{document}